\documentstyle[a4,12pt]{article}
\newcommand{\beq}{\begin{equation}}
\newcommand{\eeq}{\end{equation}}
\newcommand{\bea}{\begin{eqnarray}}
\newcommand{\eea}{\end{eqnarray}}
\newcommand{\bpi}{\begin{picture}}
\newcommand{\epi}{\end{picture}}
\newcommand{\sss}{\scriptscriptstyle}
\newcommand{\ts}{\textstyle}
\newcommand{\mb}{\mbox{}}
\newcommand{\rxb}{$\overline{R}_\xi$\ }
\newcommand{\cL}{{\cal L}}
\newcommand{\rmeff}{{\rm eff}}
\newcommand{\rmgf}{{\rm gf}}
\newcommand{\rmgh}{{\rm gh}}
\newcommand{\gh}{g\!\!\:h}
\newcommand{\half}{{\ts\frac{1}{2}}}
\newcommand{\quarter}{{\ts\frac{1}{4}}}
\newcommand{\threequarter}{{\ts\frac{3}{4}}}
\newcommand{\twothird}{{\ts\frac{2}{3}}}

\newcommand{\eighth}{{\ts\frac{1}{8}}}

\newcommand{\p}{\partial}
\newcommand{\vh}{\hat{v}}
\newcommand{\sbar}{\bar{s}}

\newcommand{\al}{\alpha}
\newcommand{\be}{\beta}
\newcommand{\ga}{\gamma}
\newcommand{\Ga}{\Gamma}
\newcommand{\de}{\delta}
\newcommand{\De}{\Delta}
\newcommand{\ep}{\epsilon}
\newcommand{\sg}{\sigma}
\newcommand{\eb}{\bar{\eta}}
\newcommand{\et}{\eta}
\newcommand{\ka}{\kappa}
\newcommand{\la}{\lambda}
\newcommand{\om}{\omega}
\newcommand{\ta}{\tau}
\newcommand{\vp}{\varphi}
\newcommand{\Th}{\Theta}
\newcommand{\Up}{\Upsilon}

\newcommand{\Da}{\De_{\sss\! A}}

\newcommand{\sa}{\sg_{\sss\!\! A}}
\newcommand{\sw}{\sg_{\sss\! W}}
\newcommand{\vs}{\vspace{0.2cm}}
\newcommand{\vsa}{\vspace{0.4cm}}
\newcommand{\vsb}{\vspace{0.8cm}}
\newcommand{\vsc}{\vspace{-4pt}}
\newcommand{\vsd}{\vspace{4pt}}
\newcommand{\invexp}{\rule{0pt}{10pt}}
\newcommand{\nn}{\nonumber\\}

\begin{document}

\begin{titlepage}
\begin{flushright} THU-93/16\\ hep-ph/9307220\\ July 1994\end{flushright}
\vsa
\begin{center}
{\large\bf A New Gauge for Computing Effective Potentials\vs\\
in Spontaneously Broken Gauge Theories\vsa\vsb\\}
Boris Kastening\vsb\\
{\it Instituut voor Theoretische Fysica\\
     Rijksuniversiteit te Utrecht\\
     Princetonplein 5\\
     P.O.\ Box 80006\\
     NL-3508 TA Utrecht\\
     The Netherlands}\vsb\vsa\\
{\bf Abstract}\\
\end{center}


A new class of renormalizable gauges is introduced that is particularly
well suited to compute effective potentials in spontaneously broken
gauge theories. It allows one to keep free gauge parameters when computing
the effective potential from vacuum graphs or tadpoles without encountering
mixed propagators of would-be-Goldstone bosons and longitudinal modes of
the gauge field. As an illustrative example several quantities are computed
within the Abelian Higgs model, which is renormalized at the two-loop level.
The zero temperature effective potential in the new gauge is compared to that
in $R_\xi$ gauge at the one-loop level and found to be not only easier to
compute but also to have a more convenient analytical structure. To
demonstrate renormalizability of the gauge for the non-Abelian case,
the renormalization of an SU(2)-Higgs model with completely broken gauge
group and of an SO(3)-Higgs model with an unbroken SO(2) subgroup is
outlined and renormalization constants are given at the one-loop level.
\end{titlepage}


\section{Introduction and Summary}

The effective potential (EP) of a relativistic quantum field theory
(QFT) \cite{GoSa..} is a useful tool for investigating several
questions of physical interest such as the vacuum structure of
the theory, inflationary cosmology and finite temperature phase
transitions (see e.g.\ \cite{KoTu..} and references therein). Recently
for instance there has been renewed interest in the details of the
electroweak phase transition in the early universe since its nature is
crucial in deciding whether the baryon asymmetry of the universe was created
at that time. The availability of a good approximation scheme for the EP
of the Higgs field is considered to be critical in that context by many
authors (see e.g.\ \cite{CoKa..} and references therein).

Although the EP of spontaneously broken gauge theories (SBGT) is
frequently employed, there is no convenient gauge that at the same time
\begin{list}{}{}
\item[(i)]   allows the EP to be computed from graphs with no or
             one external line (from here on to be called vacuum graphs
             and tadpoles, respectively),
\item[(ii)]  avoids awkward-to-use mixed propagators of would-be-Goldstone
             bosons and the longitudinal modes of the gauge field and
\item[(iii)] keeps at least one free gauge parameter.
\end{list}
Usually Landau gauge (i.e.\ $R_\xi$ gauge with $\xi=0$) is used
because then points (i) and (ii) are fulfilled. However in this case
there is no free gauge parameter left. Since in general the EP itself
is gauge dependent, it would be important to see if physical results
extracted from it are indeed gauge invariant.

In this paper a class of quadratic renormalizable gauges very similar
to the (linear) $R_\xi$ gauges is introduced that fulfills all of the
above requirements. The price one has to pay is the existence of two
or more gauge parameter of which a subset has to be set to unity to
avoid mixed propagators. This causes no problems as long as no use
is made of the renormalization group which would make all gauge
parameters running and again give rise to the presence of mixed
propagators.

The paper is structured as follows:
In section \ref{npoint} why and how vacuum graphs (or tadpoles) can be
used to compute the EP of a QFT is briefly reviewed.
In section \ref{newgauge} the new class of gauges is introduced for
a general SBGT. Its features are described and its BRST invariance
is written down.
Section \ref{ssbar} is devoted to BRST and anti-BRST invariance. It is
shown how the number of gauge parameters can be reduced by imposing
anti-BRST invariance additionally to BRST invariance.
In section \ref{abelianhiggs} the Abelian Higgs model is used as an
illustrative example. Its Feynman rules in the new class of gauges
are given, it is renormalized at the two-loop level and the physical
Higgs and gauge boson masses are computed at the one-loop level and
their gauge independence checked. The one-loop contribution to the EP
is compared to that in $R_\xi$ gauge and found not only to be easier
to compute but also to have a more desirable analytical structure. The
two-loop contribution to the EP is given.
To demonstrate the renormalizability of the gauge in more complicated
cases, in sections \ref{su2higgs} and \ref{so3higgs} the renormalization
of an SU(2)-Higgs model with completely broken gauge group and of an
SO(3)-Higgs model with an unbroken SO(2) subgroup is sketched with
explicit results given at the one-loop level. Since the evaluation of
the EP proceeds in close analogy to the Abelian case it is not considered
anew.

\section{The EP and 1PI n-Point Functions}
\label{npoint}

In this section the well-known connection between the EP of a QFT and
vacuum graphs \cite{CoWe,Ja}, tadpoles \cite{We,LeSc} and higher order
functions in a shifted theory is presented in a very compact way
for further reference.

The EP of a QFT is the generator for the one particle irreducible (1PI)
$n$-point functions at zero external momenta. If we expand the EP $V(\phi)$
about some point $\om$, we get
\beq
\label{epomega}
V(\phi) = -\sum_{n=0}^\infty\frac{1}{n!}\Gamma_{n}(\om,p_i=0)
(\phi-\om)^n,
\eeq
where $\Gamma_{n}(\om,p_i)$ is the 1PI $n$-point Greens function in
the modified theory where the field $\phi$ has been shifted by $\om$.
Taking the derivative of this equation with respect to $\om$ and
observing that the EP is independent of the point we expand it about,
we arrive at the recursion relation
\beq
\Gamma_{n+1}(\om,p_i=0)
=\frac{\partial}{\partial\om}\Gamma_{n}(\om,p_i=0).
\eeq
Setting $\phi=\om$ in (\ref{epomega}) gives
\beq
V(\om)=-\Gamma_{0}(\om,p_i=0)
\eeq
and therefore the $n$-th derivative of the EP is given by
\beq
V^{(n)}(\om)=-\Gamma_{n}(\om,p_i=0).
\eeq
Thus instead of summing over $n$ as in (\ref{epomega}) we can
use $n$-point functions in a shifted theory and integrate $n$ times.
However, if higher order functions than tadpoles are used,
the choice of integration constants is a non-trivial problem
if no additional information is available.
Throughout this paper vacuum graphs (i.e.\ $n=0$) are used so that
no integration over $\om$ is needed.

It is crucial for the above derivation to hold that the only place where
$\om$ enters the modified Lagrangian is through the shifted scalar field
as already noted in \cite{We}. Otherwise (\ref{epomega}) does not hold.

\section{The New Gauge}
\label{newgauge}

Within the scalar sector of a gauge field theory with gauge group G let
$\Phi$ be a field that is expected to get a vacuum expectation value
(vev) so that G is spontaneously broken. Assume that $\Phi$ has been
put into a real multiplet on which a homogeneous linear and unitary
representation acts and that, for simplicity, G is simple and the
representation irreducible. (The generalization to non-simple groups
and reducible representations is straightforward.)

First recall the class of $R_\xi$ gauges \cite{FuLe..}:
The gauge fixing term here is
\beq
\label{rxi}
\cL_\rmgf= -\frac{1}{2\xi}\left(\partial_\mu A_a^\mu
+i\xi g\;v_0^TT_a\Phi'\right)^2,
\eeq
where $\Phi=v_0+\Phi'$, $v_0$ being the tree-level vev of $\Phi$,
$g$ the gauge coupling and $T_a$ the (imaginary and antisymmetric)
gauge group generators. With this gauge fixing it is not possible to
satisfy all three requirements (i)-(iii) simultaneously: If the Higgs
field is shifted by some $\om$ (additionally to $v_0$), then point (ii)
is violated. If we modify $v_0^T$ in the gauge fixing term so as to avoid
that, then the complete Lagrangian does no longer depend only on the
sum of shifted field and shift, but on those two quantities separately
which invalidates the derivation in section \ref{npoint} and point (i)
is no longer fulfilled. If we set $\xi=0$, it turns out that (i) and (ii)
are satisfied (most easily seen by the fact that the gauge to be introduced
subsequently has effectively the same Feynman rules in the corresponding
limit), but (iii) is violated, i.e.\ we have no check of gauge independence
of physical quantities anymore.

Now choose a unit vector $\vh$ in $\Phi$ space and consider the gauge
fixing term
\beq
\label{rxibar}
\cL_\rmgf=-\half\sg_{ab}F_aF_b\,
\eeq
with
\beq
\label{Fa}
F_a=\p_\mu A_a^\mu+ig\Th_{ab}\phi^T\vh\vh^TT_b\phi+gR_{abc}\eb_b\et_c
+gS_{abc}A_{b\mu}A_c^{\mu}\,,
\eeq
where $\sg_{ab}$ and $\Th_{ab}$ are real and symmetric, $\eb_a,\et_a$
are anti-commuting ghost fields and repeated Latin indices are summed
over all generator indices of the gauge group. This term necessitates
the further addition of the ghost term
\bea
\label{ghost}
\cL_\rmgh&=&(\p_\mu\eb_a)(D^\mu\et_a)
-g^2\Th_{ac}\phi^T(\vh\vh^TT_cT_b-T_b\vh\vh^TT_c)\phi\eb_a\et_b\nn
&&\mb+\quarter g^2R_{abe}f_{ecd}\eb_a\eb_b\et_c\et_d
-2gS_{abc}\eb_aA_b^\mu D_\mu\et_c
\eea
with
\beq
D^\mu\et_a=\partial^\mu\et_a+gf_{abc}\et_bA_c^\mu,
\eeq
where the $f_{abc}$ are the completely antisymmetric structure
constants of our gauge group with normalization $[T_a,T_b]=if_{abc}T_c$.

The action is now invariant under the BRST transformation
\beq
\begin{array}{lcl}
\de\phi&=&-i\ka g\et_aT_a\phi\\
\de A_a^\mu&=&\ka D^\mu\et_a\\
\de\eb_a&=&-\ka\sg_{ab}F_b\\
\de\et_a&=&\half\ka gf_{abc}\et_b\et_c
\end{array}
\eeq
with anticommuting $\ka$, provided $R_{abc}$ is antisymmetric in its first
two indices and $S_{abc}$ is symmetric in its last two indices, i.e.\
\beq
\label{rssym}
R_{abc}=-R_{bac}\,,\;\;\;\;\;\;\;\;\;\;\;\;S_{abc}=S_{acb}\,.
\eeq

Note that the terms ``gauge fixing'' and ``ghost'' Lagrangian are
somewhat ambiguous since $\cL_\rmgf$ already contains the
ghost fields. This, as well as the appearance of quartic ghost
terms, is quite generic to quadratic gauge fixing functions $F_a$
\cite{quadgf} and requires us to BRST-quantize \cite{BRST} the
theory instead of following the Fadeev-Popov procedure.
Note that if we ignore for a moment the quadratic gauge and ghost
terms in the gauge fixing function (\ref{Fa}) and set
$\Th_{ab}=\sg_{ab}=\de_{ab}$, we could loosely say that the new gauge
obtains from $R_\xi$ gauge by promoting the quantity $\vh_0^T$ in
(\ref{rxi}) to a full field including its quantum fluctuations.

We assume here that it is possible to choose the $\sg_{ab}$, $\Th_{ab}$,
$R_{abc}$ and $S_{abc}$ in such a way that renormalization forces no new
terms in the Lagrangian upon us. If the $\vh^TT_aT_b\Phi$ cannot be
expressed by $\vh^T\Phi$ and the $\vh^TT_a\Phi$ that might not be true
and additional terms like e.g.\ $\Up_{abcd}\Phi^TT_b\vh\vh^TT_cT_d\Phi$
have to be introduced into the gauge fixing function. Instead of trying
to set up a general procedure to pick the $\sg_{ab}$, $\Th_{ab}$, $R_{abc}$,
$S_{abc}$, $\Up_{abcd}$, \ldots, examples will be given in sections
\ref{abelianhiggs}, \ref{su2higgs} and \ref{so3higgs}. See however the
last paragraph of section \ref{ssbar}.

Now let us introduce some constant shift $\vp$ by
\beq
\label{shift}
\Phi=\vp\vh+\Phi'.
\eeq
Then $\cL_\rmgf$ and $\cL_\rmgh$
still depend only on the sum of shift and remaining
quantum field $\Phi'$ and the derivation in section \ref{epomega} is
valid and thus point (i) is fulfilled.

On the other hand, it is easy to check that for any shift $\vp$
the mixing terms between would-be-Goldstone modes and longitudinal
gauge modes vanish and point (ii) is also fulfilled, provided
$(\sg\Th)_{ab}=\de_{ab}$ when $a$ or $b$ corresponds to a broken
generator, i.e.\ $T_{a,b}\vh\neq 0$. Although $(\sg\Th)_{ab}$
generally gets renormalized, we can set its renormalized value among
the broken generators equal to $\de_{ab}$ (in BPHZ renormalization)
and point (ii) remains valid.

Point (iii) is clearly satisfied, too, since $\sg_{ab}$ and $\Th_{ab}$
contain gauge parameters.

Replacing $(\sg,\Th,R,S)\rightarrow(\sg/\xi,\xi\Th,\xi R,\xi S)$ and
letting $\xi\rightarrow 0$ one regains effectively the Feynman rules
for Landau gauge among the $R_\xi$ gauges. This explains naturally why
in this gauge vacuum graphs or tadpoles can be used to compute
the EP. However it is not clear to the present author if also the
effective potential will always go smoothly to that of Landau gauge if the
limit $\xi\rightarrow 0$ is taken after the regulator is removed.
This will only happen if even for $\vp\neq v_0$ all infrared singularities
stemming from vanishing ghost and longitudinal gauge boson masses cancel.
It is easily seen, however, that this problem does not appear up to two
loops.

Setting $(\sg\Th)_{ab}=\de_{ab}$ for broken generators is unproblematic
though, since no eigenvalues of the mass matrix disappear and therefore
there is no discontinuos change in the singularity structure of the theory
to any given order in the loop expansion.

It is important to ensure that $F_a$ does not receive an expectation value
since otherwise BRST symmetry would be spontaneously broken and the gauge
a bad one in the sense of \cite{dWPa}. As in other gauges, the easiest way
to achieve this is to impose appropriate symmetries on $\cL_\rmeff$ as is
the case in all examples in this paper.

The class of gauges introduced by (\ref{rxibar}) is quadratic and
renormalizable for appropriate $\sg_{ab}$, $\Th_{ab}$, $R_{abc}$ and
$S_{abc}$ (up to the issue of closing the algebra of scalar fields appearing
in $\cL_\rmgf$ and $\cL_\rmgh$, see above and next section). Because of its
similarity to the class of $R_\xi$ gauges, it is called \rxb from here on.

\section{BRST and anti-BRST Invariance}
\label{ssbar}

In this section the gauge fixing and ghost Lagrangians (\ref{rxibar}) and
(\ref{ghost}) will be derived in a simple way. Then a similar procedure
will be followed to find gauge fixing and ghost Lagrangians that are
additionally anti-BRST invariant.

Define nil-potent BRST and anti-BRST operators $s$ and $\sbar$ by
\cite{OjBa..}
\beq
\label{brst}
\begin{array}{lcllcl}
s\phi&=&-ig\et_aT_a\phi&\sbar\phi&=&-ig\eb_aT_a\phi\vsd\\
sA_a^\mu&=&D^\mu\et_a&\sbar A_a^\mu&=&D^\mu\eb_a\vsd\\
s\et_a&=&\half gf_{abc}\et_b\et_c\;\;\;\;\;\;\;\;
&\sbar\eb_a&=&\half gf_{abc}\eb_b\eb_c\vsd\\
s\eb_a&=&B_a&\sbar\et_a&=&-B_a+gf_{abc}\eb_b\et_c\vsd\\
sB_a&=&0&\sbar B_a&=&gf_{abc}\eb_bB_c
\end{array}
\eeq
with
\bea
D^\mu\et_a&=&\p^\mu\et_a+gf_{abc}\et_bA_c^\mu\,,\\
D^\mu\eb_a&=&\p^\mu\eb_a+gf_{abc}\eb_bA_c^\mu\,,
\eea
and where $B_a$ are auxiliary fields, i.e.\ without kinetic term. Consider
\bea
\cL_{\rm gf+gh}
&=&s[\eb_a(\p_\mu A_a^\mu+ig\Th_{ab}\phi^T\vh\vh^TT_b\phi
+\half gR_{abc}\eb_b\et_c+gS_{abc}A_{b\mu}A_c^\mu
+\half\sg_{ab}^{-1}B_b)]\nn
&=&\half\sg_{ab}^{-1}B_aB_b\nn
&&\mb+B_a[\p_\mu A_a^\mu+ig\Th_{ab}\phi^T\vh\vh^TT_b\phi
+\half g(R_{abc}-R_{bac})\eb_b\et_c+gS_{abc}A_{b\mu}A_c^\mu]\nn
&&\mb+\p_\mu\eb_aD^\mu\et_a
-g^2\Th_{ac}\phi^T(\vh\vh^TT_cT_b-T_b\vh\vh^TT_c)\phi\eb_a\et_b\nn
&&\mb+\quarter g^2R_{abe}f_{ecd}\eb_a\eb_b\et_c\et_d
-g(S_{abc}+S_{acb})\eb_aA_{b\mu}D^\mu\et_c
\eea
with some real quantities $\sg_{ab}$, $\Th_{ab}$, $R_{abc}$ and
$S_{abc}$. Clearly we can assume $R_{abc}$ to be anti-symmetric
in its first two and $S_{abc}$ to be symmetric in its last two indices.
Nil-potency of $s$ makes $\cL_{\rm gf+gh}$ automatically $s$-invariant.
Integrating out $B_a$ of the generating functional or, equivalently,
eliminating it from the Lagrangian by its equation of motion, one recovers
the sum of (\ref{rxibar}) and (\ref{ghost}).

Now consider the $s$- and $\sbar$-invariant gauge fixing and ghost Lagrangian
\beq
\cL_{s\sbar}
=\half s\sbar(\om\phi^T\vh\vh^T\phi+\Ga_{ab}\eb_a\et_b
+\om\De^{-1}_{ab}A_{a\mu}A_b^\mu)
\eeq
with real $\om$ and real and symmetric $\Ga_{ab}$ and $\De_{ab}$.
Applying the rules for $s$ and $\sbar$, integrating out $B_a$,
rescaling $\eb_a\rightarrow-\om^{-1}\De_{ab}\eb_b$ and defining
$\sg_{ab}=\om^2(\De\Ga\De)_{ab}^{-1}$, one gets
\bea
\cL_{s\sbar}
&=&-\half\sg_{ab}F_aF_b+(\p_\mu\eb_a)(D^\mu\et_a)
-g^2\De_{ac}\phi^T(\vh\vh^TT_cT_b-T_b\vh\vh^TT_c)\phi\eb_a\et_b\nn
&&\mb+\quarter g^2R_{abe}f_{ecd}\eb_a\eb_b\et_c\et_d
-2gS_{abc}\eb_aA_{b\mu}D^\mu\et_c
\eea
with
\bea
F_a&=&\p_\mu A_a^\mu+ig\De_{ab}\phi^T\vh\vh^TT_b\phi+gR_{abc}\eb_b\et_c
+gS_{abc}A_{b\mu}A_c^{\mu}\,,\\
\label{rabc}
R_{abc}
&=&\half[\De_{ad}\De_{be}\De_{cf}^{-1}\sg_{fg}^{-1}\De_{gh}^{-1}f_{deh}
-(\sg_{ad}^{-1}\De_{bf}-\sg_{bd}^{-1}\De_{af})\De_{de}^{-1}f_{cef}],\\
\label{sabc}
S_{abc}
&=&-\half\De_{ae}(f_{ebd}\De_{dc}^{-1}+f_{ecd}\De_{db}^{-1}).
\eea
Thus, given $\cL_\rmgf$ and $\cL_\rmgh$ by
(\ref{rxibar})-(\ref{ghost}), $\cL_\rmgf+\cL_\rmgh$ is not
only BRST but also anti-BRST invariant, if there exists $\De_{ab}$ such that
\beq
\label{theta}
\Th_{ab}=\De_{ab}\;\;\;{\rm for}\;\;\;T_b\vh\neq 0,
\eeq
i.e.\ for $b$ corresponding to broken generators, and such that $R_{abc}$
and $S_{abc}$ fulfill (\ref{rabc}) and (\ref{sabc}). In the examples
of sections \ref{su2higgs} and \ref{so3higgs} we will check at the one-loop
level that the resulting conditions on the gauge parameters are stable
under renormalization.

There is a subtlety about unbroken Abelian subgroups. In this case it
can be consistent to have an invertible $\sg_{ab}$ but to set $\De_{ab}$
and $\De_{ab}^{-1}$ to zero when $a,b$ refer to that subgroup such that
they are their mutually inverses only for the restriction to the rest
of indices. It is easily seen that then the ghost fields corresponding to
the Abelian subgroup under consideration effectively drop from the theory.
An example of this case is the SO(3)-Higgs model with its unbroken
SO(2) subgroup discussed in section \ref{so3higgs}.

Again it needs to be noted that renormalization might force new terms
in the Lagrangian on us, if $\vh^TT_aT_b\Phi$ cannot be expressed
through $\vh^T\Phi$ and the $\vh^TT_a\Phi$. Then one can try to introduce
additional terms like e.g.\ $s\sbar(\Up_{ab}\Phi^TT_a\vh\vh^TT_b\Phi)$ into
$\cL_{s\sbar}$ until the algebra of scalar fields appearing in $\cL_{s\sbar}$
closes.

To close this section we can sketch a method of obtaining the desired
renormalizable effective Lagrangian even when renormalization forces
terms upon us that are not taken care of by (\ref{rxibar})-(\ref{ghost}):
Given some field content (including ghost and auxiliary fields)
as well as gauge group generators and structure constants, write down the
most general ${\rm dim}\leq 4$ Lagrangian invariant under the $s$-symmetry
of (\ref{brst}) (possibly extended to include more fields, e.g.\ fermions).
Impose other global symmetries (discrete and/or continous, e.g.\ anti-BRST)
to prevent the resulting $F_a$ from getting an expectation value,
to possibly eliminate other unwanted terms and to diminish the number of
gauge parameters. However keep $\cL_\rmeff$ general enough such
that when computing the effective potential a suitable subset of gauge
parameters can be adjusted to the effect that unwanted mixings of fields
disappear and still at least one gauge parameter is left.

\section{Application I: The Abelian Higgs Model}
\label{abelianhiggs}

In this section the use of \rxb gauge is illustrated
within the Abelian Higgs model. It is given by the Lagrange density
\beq
\label{abhi}
\cL = \frac{1}{2}\left(D_\mu\Phi\right)^T\left(D^\mu\Phi\right)
-\frac{1}{4}F_{\mu\nu}F^{\mu\nu}-\frac{1}{2}m^2\Phi^T\Phi
-\frac{\la}{4}(\Phi^T\Phi)^2
\eeq
with
\bea
F_{\mu\nu}&=&\partial_\mu A_\nu-\partial_\nu A_\mu\,,
\\
D_\mu\Phi&=&\left(\partial_\mu+igT_1A_\mu\right)\Phi\,,
\\
T_1&=&\ta_2
=(\raisebox{-3pt}{$\stackrel{\scriptstyle 0}{\scriptstyle i}
\;\stackrel{\scriptstyle -i}{\scriptstyle 0}$})\,,
\\
\Phi^T&=&(\phi_1,\phi_2)\,.
\eea
Here $\phi_1$ and $\phi_2$ are real scalar fields and with the scalar
self-coupling $\la>0$ and $m^2<0$ the U(1) gauge symmetry is spontaneously
broken. $s$-invariance forces $R_{abc}=0$. Imposing the symmetries
$(A_\mu,\phi_1)\rightarrow(-A_\mu,-\phi_1)$ and
$(A_\mu,\phi_2)\rightarrow(-A_\mu,-\phi_2)$ of $\cal L$ also on
$\cL_\rmgf+\cL_\rmgh$, we get $S_{abc}=0$.

If we choose $\vh^T=(1,0)$, $\sg_{11}=\sg/\xi$ and $\Th_{11}=\xi$ we have
\bea
\label{abhigf}
\cL_\rmgf
&=&
-\frac{\sg}{2\xi}\left(\partial_\mu A^\mu
+\xi g\phi_1\phi_2\right)^2,
\\
\label{abhigh}
\cL_\rmgh
&=&
\partial_\mu\eb\partial^\mu\et
-\xi g^2\left(\phi_1^2-\phi_2^2\right)\eb\et.
\eea

$\cL_\rmeff=\cL+\cL_\rmgf+\cL_\rmgh$ can also be obtained by the following
procedure: Take the most general ${\rm dim}\leq 4$ Lagrangian with given
field content $(\phi_1,\phi_2,A_\mu,\eb,\et,B)$. Impose the nil-potent
$s$-symmetry of (\ref{brst}) with  $T_1$ given above and $f_{abc}=0$ and
the discrete symmetries $(A_\mu,\phi_1,B) \rightarrow (-A_\mu,-\phi_1,-B)$
and $(A_\mu,\phi_2,B) \rightarrow (-A_\mu,-\phi_2,-B)$. Integrate out $B$.
Up to total divergencies and trivial changes of variables, $\cL_\rmeff$
with the parameters given above is the result. Therefore $\cL_\rmeff$ is
renormalizable in any regularization scheme that observes the symmetries,
e.g.\ dimensional regularization.

Since $\cL_\rmeff$ is now invariant under
$(\eb,\et)\rightarrow(-\et,\eb)$ and since $f_{abc}=0$, and thus this
operation relates $s$- and $\sbar$-invariance, $\cL_\rmeff$ is
also $\sbar$-invariant.

Now we are ready to break the U(1) symmetry by shifting
$(\phi_1,\phi_2) = (\vp+\phi_1',\phi_2')$. For $\sg=1$,
there is no mixing between would-be-Goldstone and longitudinal gauge field
modes and $\cL_\rmeff$ becomes the sum of the following four pieces, ordered
by their dimension:
\bea
\cL_0 &=&
-\frac{1}{4}\la\vp^4-\frac{1}{2}m^2\vp^2,\\
\cL_1 &=&
-(\la\vp^3+m^2\vp)\phi_1',\\
\cL_2 &=&
\frac{1}{2}(\partial_\mu\phi_1')^2-\frac{1}{2}(3\la\vp^2+m^2)\phi_1'^2
+\frac{1}{2}(\partial_\mu\phi_2')^2
-\frac{1}{2}[(\la+\xi g^2)\vp^2+m^2]\phi_2'^2\nn
&&-\frac{1}{4}F_{\mu\nu}F^{\mu\nu}-\frac{1}{2\xi}(\partial_\mu A^\mu)^2
+\frac{1}{2}g^2\vp^2A_\mu A^\mu+\partial_\mu\eb\partial^\mu\et
-\xi g^2\vp^2\eb\et,\\
\cL_{3,4} &=&
-\frac{1}{4}\la\phi_1'^4-\la\vp\phi_1'^3
-\frac{1}{2}(\la+\xi g^2)\phi_1'^2\phi_2'^2
-(\la+\xi g^2)\vp\phi_1'\phi_2'^2
-\frac{1}{4}\la\phi_2'^4\nn
&&+g^2\vp\phi_1'A_\mu A^\mu+2g\phi_2'A_\mu\partial^\mu\phi_1'
+\frac{1}{2}g^2\phi_1'^2A_\mu A^\mu
+\frac{1}{2}g^2\phi_2'^2A_\mu A^\mu\nn
&&-2\xi g^2\vp\phi_1'\eb\et
-\xi g^2\phi_1'^2\eb\et
+\xi g^2\phi_2'^2\eb\et.
\eea
The Feynman rules can immediately be read off and are given in
table \ref{feynmanrules}.
\begin{table}[t]
\begin{tabular}{lc|c|c}
&&\rxb&$R_\xi$\\
\hline
\underline{\bf constant:}
&\bpi(0,22)\put(0,3){\circle*{3}}\epi
&$\displaystyle
-i\left(\frac{1}{4}\la\vp^4+\frac{1}{2}m^2\vp^2\right)$&same\\
\underline{\bf tadpole:}
&
\bpi(0,18)
\put(0,10){\circle*{3}}\put(0,-5){\line(0,1){15}}
\epi
&$-i(\la\vp^3+m^2\vp)$&same\\
&&&\\
\underline{\bf propagators:}&&&\vsc\vsc\\
Higgs:&
\bpi(36,25)
\put(0,3){\line(1,0){36}}
\epi
&$\displaystyle\frac{i}{k^2-m_H^2}$&same\vsc\\
\protect\parbox{55pt}{would-be-\\Goldstone:}
&
\bpi(36,25)
\multiput(0,3)(8,0){5}{\line(1,0){4}}
\epi
&$\displaystyle\frac{i}{k^2-m_G^2}$&same\vsc\\
gauge:&
\bpi(48,30)(-8,0)
\put(-9,0){$\mu$}
\put(34,0){$\nu$}
\multiput(2,3)(8,0){4}{\oval(4,4)[t]}
\multiput(6,3)(8,0){4}{\oval(4,4)[b]}
\epi
&$\displaystyle
-i\left[\frac{g_{\mu\nu}-k_\mu k_\nu/k^2}{k^2-m_A^2}
+\frac{\xi k_\mu k_\nu/k^2}{k^2-m_{\gh}^2}\right]$&same\vsc\\
ghost:&
\bpi(36,25)
\multiput(0,3)(4,0){10}{\circle*{1}}
\multiput(18.5,3)(-0.1,0.1){30}{\line(1,0){0.1}}
\multiput(18.5,3)(-0.1,-0.1){30}{\line(1,0){0.1}}
\epi
&$\displaystyle\frac{i}{k^2-m_{\gh}^2}$&same\\
\end{tabular}
\protect\vsa\\
\begin{tabular}{c|c|c||c|c|c}
\underline{\bf vertices:}&\rxb&$R_\xi$&
\underline{\bf vertices:}&\rxb&$R_\xi$\\
\hline

\bpi(32,40)
\put(0,0){\line(1,1){16}}
\put(16,16){\line(1,-1){16}}
\put(16,16){\line(0,1){20}}
\put(16,16){\circle*{3}}
\epi
&\raisebox{12pt}{$-6i\la\vp$}&\raisebox{12pt}{same}&
\bpi(32,40)
\multiput(0,2)(4,4){4}{\oval(4,4)[br]}
\multiput(4,2)(4,4){4}{\oval(4,4)[tl]}
\multiput(16,14)(4,-4){4}{\oval(4,4)[tr]}
\multiput(20,14)(4,-4){4}{\oval(4,4)[bl]}
\put(16,16){\line(0,1){20}}
\put(16,16){\circle*{3}}
\put(-8,-2){$\mu$}
\put(34,-2){$\nu$}
\epi
&\raisebox{12pt}{$2ig^2\vp g_{\mu\nu}$}&\raisebox{12pt}{same}\\
\bpi(32,36)
\put(0,0){\line(1,1){32}}
\put(0,32){\line(1,-1){32}}
\put(16,16){\circle*{3}}
\epi
&\raisebox{12pt}{$-6i\la$}&\raisebox{12pt}{same}&
\bpi(32,36)
\multiput(0,2)(4,4){4}{\oval(4,4)[br]}
\multiput(4,2)(4,4){4}{\oval(4,4)[tl]}
\multiput(16,14)(4,-4){4}{\oval(4,4)[tr]}
\multiput(20,14)(4,-4){4}{\oval(4,4)[bl]}
\put(16,16){\line(1,1){16}}
\put(0,32){\line(1,-1){16}}
\put(16,16){\circle*{3}}
\put(-8,-2){$\mu$}
\put(34,-2){$\nu$}
\epi
&\raisebox{12pt}{$2ig^2g_{\mu\nu}$}&\raisebox{12pt}{same}\\
\bpi(32,36)
\multiput(0,0)(0.25,0.25){12}{\line(1,0){0.1}}
\multiput(6,6)(0.25,0.25){12}{\line(1,0){0.1}}
\multiput(12,12)(0.25,0.25){12}{\line(1,0){0.1}}
\multiput(17,17)(0.25,0.25){12}{\line(1,0){0.1}}
\multiput(23,23)(0.25,0.25){12}{\line(1,0){0.1}}
\multiput(29,29)(0.25,0.25){12}{\line(1,0){0.1}}
\multiput(0,32)(0.25,-0.25){12}{\line(1,0){0.1}}
\multiput(6,26)(0.25,-0.25){12}{\line(1,0){0.1}}
\multiput(12,20)(0.25,-0.25){12}{\line(1,0){0.1}}
\multiput(17,15)(0.25,-0.25){12}{\line(1,0){0.1}}
\multiput(23,9)(0.25,-0.25){12}{\line(1,0){0.1}}
\multiput(29,3)(0.25,-0.25){12}{\line(1,0){0.1}}
\put(16,16){\circle*{3}}
\epi
&\raisebox{12pt}{$-6i\la$}&\raisebox{12pt}{same}&
\bpi(32,36)
\multiput(0,2)(4,4){4}{\oval(4,4)[br]}
\multiput(4,2)(4,4){4}{\oval(4,4)[tl]}
\multiput(16,14)(4,-4){4}{\oval(4,4)[tr]}
\multiput(20,14)(4,-4){4}{\oval(4,4)[bl]}
\multiput(17,17)(0.25,0.25){12}{\line(1,0){0.1}}
\multiput(23,23)(0.25,0.25){12}{\line(1,0){0.1}}
\multiput(29,29)(0.25,0.25){12}{\line(1,0){0.1}}
\multiput(0,32)(0.25,-0.25){12}{\line(1,0){0.1}}
\multiput(6,26)(0.25,-0.25){12}{\line(1,0){0.1}}
\multiput(12,20)(0.25,-0.25){12}{\line(1,0){0.1}}
\put(16,16){\circle*{3}}
\put(-8,-2){$\mu$}
\put(34,-2){$\nu$}
\epi
&\raisebox{12pt}{$2ig^2g_{\mu\nu}$}&\raisebox{12pt}{same}\\
\bpi(32,38)
\multiput(0,0)(0.25,0.25){12}{\line(1,0){0.1}}
\multiput(6,6)(0.25,0.25){12}{\line(1,0){0.1}}
\multiput(12,12)(0.25,0.25){12}{\line(1,0){0.1}}
\multiput(17,15)(0.25,-0.25){12}{\line(1,0){0.1}}
\multiput(23,9)(0.25,-0.25){12}{\line(1,0){0.1}}
\multiput(29,3)(0.25,-0.25){12}{\line(1,0){0.1}}
\put(16,16){\line(0,1){20}}
\put(16,16){\circle*{3}}
\epi
&\raisebox{12pt}{$-2i(\la+\xi g^2)\vp$}
&\raisebox{12pt}{$-2i\la\vp$}&
\bpi(32,38)
\multiput(0,0)(3,3){6}{\circle*{1}}
\multiput(32,0)(-3,3){6}{\circle*{1}}
\put(16,16){\line(0,1){20}}
\put(23.5,4.5){\line(1,0){4}}
\put(27.5,4.5){\line(0,1){4}}
\put(3.5,7.5){\line(1,0){4}}
\put(7.5,3.5){\line(0,1){4}}
\put(16,16){\circle*{3}}
\epi
&\raisebox{12pt}{$-2i\xi g^2\vp$}&\raisebox{12pt}{$-i\xi g^2\vp$}\\
\bpi(32,36)
\multiput(0,0)(0.25,0.25){12}{\line(1,0){0.1}}
\multiput(6,6)(0.25,0.25){12}{\line(1,0){0.1}}
\multiput(12,12)(0.25,0.25){12}{\line(1,0){0.1}}
\multiput(17,15)(0.25,-0.25){12}{\line(1,0){0.1}}
\multiput(23,9)(0.25,-0.25){12}{\line(1,0){0.1}}
\multiput(29,3)(0.25,-0.25){12}{\line(1,0){0.1}}
\put(16,16){\line(1,1){16}}
\put(0,32){\line(1,-1){16}}
\put(16,16){\circle*{3}}
\epi
&\raisebox{12pt}{$-2i(\la+\xi g^2)$}&\raisebox{12pt}{$-2i\la$}&
\bpi(32,36)
\multiput(0,0)(3,3){6}{\circle*{1}}
\multiput(32,0)(-3,3){6}{\circle*{1}}
\put(16,16){\line(1,1){16}}
\put(0,32){\line(1,-1){16}}
\put(23.5,4.5){\line(1,0){4}}
\put(27.5,4.5){\line(0,1){4}}
\put(3.5,7.5){\line(1,0){4}}
\put(7.5,3.5){\line(0,1){4}}
\put(16,16){\circle*{3}}
\epi
&\raisebox{12pt}{$-2i\xi g^2$}&\raisebox{12pt}{---}\\
\bpi(32,40)
\put(0,0){\line(1,1){16}}
\multiput(17,15)(0.25,-0.25){12}{\line(1,0){0.1}}
\multiput(23,9)(0.25,-0.25){12}{\line(1,0){0.1}}
\multiput(29,3)(0.25,-0.25){12}{\line(1,0){0.1}}
\multiput(16,18)(0,8){3}{\oval(4,4)[l]}
\multiput(16,22)(0,8){2}{\oval(4,4)[r]}
\put(22,6){\line(1,0){4}}
\put(26,6){\line(0,1){4}}
\put(5,9){\line(1,0){4}}
\put(9,5){\line(0,1){4}}
\put(16,16){\circle*{3}}
\put(-6,9){$k_1$}
\put(28,9){$k_2$}
\put(20,34){$\mu$}
\epi
&\raisebox{12pt}{$2gk_{1\mu}$}&\raisebox{12pt}{$g(k_{1}+k_{2})_\mu$}&
\bpi(32,40)
\multiput(17,17)(0.25,0.25){12}{\line(1,0){0.1}}
\multiput(23,23)(0.25,0.25){12}{\line(1,0){0.1}}
\multiput(29,29)(0.25,0.25){12}{\line(1,0){0.1}}
\multiput(0,32)(0.25,-0.25){12}{\line(1,0){0.1}}
\multiput(6,26)(0.25,-0.25){12}{\line(1,0){0.1}}
\multiput(12,20)(0.25,-0.25){12}{\line(1,0){0.1}}
\put(16,16){\circle*{3}}
\multiput(0,0)(3,3){6}{\circle*{1}}
\multiput(32,0)(-3,3){6}{\circle*{1}}
\put(23.5,4.5){\line(1,0){4}}
\put(27.5,4.5){\line(0,1){4}}
\put(3.5,7.5){\line(1,0){4}}
\put(7.5,3.5){\line(0,1){4}}
\epi
&\raisebox{12pt}{$2i\xi g^2$}&\raisebox{12pt}{---}
\end{tabular}
\caption{Feynman rules for the Abelian Higgs model in \rxb gauge for
$\sg_R=1$ and in generalized $R_\xi$ gauge with $m_H^2=3\la\vp^2+m^2$,
$m_G^2=(\la+\xi g^2)\vp^2+m^2$, $m_A^2=g^2\vp^2$, $m_{\gh}^2=\xi g^2\vp^2$.
$k_\mu$ is the momentum flowing through the propagators.}
\label{feynmanrules}
\end{table}
For comparison also the Feynman rules in generalized $R_\xi$ gauge, i.e.\
\bea
\cL_\rmgf=-1/(2\xi)(\partial_\mu A^\mu+\xi g\vp\phi_2')^2,\\
\cL_\rmgh=\partial_\mu\eb\partial^\mu\et-\xi g^2\vp(\phi_1'+\vp)\eb\et,
\eea
are given (`generalized' since the gauge parameter $\vp$ is not necessarily
the vev of the Higgs field). Note that as promised in Landau gauge the
Feynman rules of both classes of gauges become identical (the remaining
difference in the Higgs-gauge-Goldstone vertex is immediately seen to be
irrelevant because now the gauge propagator is transverse) and therefore
$\vp$ in generalized $R_\xi$ Landau gauge can serve as the argument of the
EP. It can be shown that the Feynman rules effectively coincide also for
unitary gauge ($\xi\rightarrow\infty$) in both classes of gauges.

Up to an unphysical constant term the effective Lagrange density can be
renormalized as
\bea
\cL_\rmeff
\!\!\!&=&\!\!\!
\frac{1}{2}(\partial_\mu\phi_{1B}+g_BA_{B\mu}\phi_{2B})^2
+\frac{1}{2}(\partial_\mu\phi_{2B}-g_BA_{B\mu}\phi_{1B})^2
-\frac{1}{4}F_{B\mu\nu}F_B^{\mu\nu}
\nn
&&\!\!\!
-\frac{1}{2}m_B^2(\phi_{1B}^2+\phi_{2B}^2)
-\frac{1}{4}\la_B(\phi_{1B}^2+\phi_{2B}^2)^2
-\frac{\sg_B}{2\xi_B}(\partial_\mu A_B^\mu+\xi_B g_B\phi_{1B}\phi_{2B})^2
\nn
&&\!\!\!
+\partial_\mu\eb_B\partial^\mu\et_B-\xi_B g_B^2(\phi_{1B}^2
-\phi_{2B}^2)\eb_B\et_B
\eea
with
\beq
\label{zx}
\begin{array}{c}
\phi_{1B}=Z_H^\frac{1}{2}\phi_{1R}\,,\;\;\;
\phi_{2B}=Z_G^\frac{1}{2}\phi_{2R}\,,\;\;\;
A_{B\mu}=Z_A^\frac{1}{2}A_{R\mu}\,,\;\;\;
\eb_B\et_B=Z_\et\eb_R\et_R\,,\\
m_B=Z_m^\frac{1}{2}m_R\,,\;\;\;
\la_B=Z_\la\la_R\,,\;\;\;
g_B=Z_g^\frac{1}{2}g_R\,,\;\;\;
\xi_B=Z_\xi\xi_R\,,\;\;\;
\sg_B=Z_\sg\sg_R\,,
\end{array}
\eeq
where renormalization constants $Z_x$ have been introduced and ``$B$'' and
``$R$'' denote bare and renormalized quantities, respectively. The Ward
identity requires $Z_gZ_A=1$. Using dimensional regularization \cite{tHVe..}
and the MS scheme \cite{tH} (as for all calculations
in this paper) I have computed the $Z_x$ at the two-loop level for
$\sg_R=1$. The result is given in the appendix.

As a check on the consistency of the gauge (\ref{rxibar}) and as an
illustration for its use, the one-loop radiative corrections to the Higgs
and physical gauge boson masses are computed in the next two paragraphs.

With the tree-level Higgs field vev $v_0^2=-m^2/\la$ the tree-level
Higgs mass obtains as $m_{H_0}^2=3\la v_0^2+m^2=-2m^2$. Its one-loop
correction is given by \cite{ApCa..} (let $k_\mu$ be the momentum flowing
through the graphs)
\beq
m_{H_1}^2=i\left.\left[
\bpi(72,26)(-4,12)
\put(0,16){\line(1,0){16}}
\put(48,16){\line(1,0){16}}
\put(32,16){\circle{32}}
\put(16,16){\circle*{3}}
\put(48,16){\circle*{3}}
\put(32,17){\makebox(0,0)[b]{\footnotesize 1-loop}}
\put(32,13){\makebox(0,0)[t]{\mbox{\footnotesize 1PI}}}
\epi
+
\bpi(48,26)(-8,16.5)
\put(0,0){\line(1,0){32}}
\put(16,0){\line(0,1){8}}
\put(16,24){\circle{32}}
\put(16,0){\circle*{3}}
\put(16,8){\circle*{3}}
\put(16,25){\makebox(0,0)[b]{\footnotesize 1-loop}}
\put(16,21){\makebox(0,0)[t]{\mbox{\footnotesize 1PI}}}
\epi
\right]\right|_{\scriptstyle k^2=m_{H_0}^2,\;\vp^2=v_0^2}\;,
\eeq
where in \rxb gauge
\bea
\bpi(40,26)(28,13)
\put(0,16){\line(1,0){16}}
\put(48,16){\line(1,0){16}}
\put(32,16){\circle{32}}
\put(16,16){\circle*{3}}
\put(48,16){\circle*{3}}
\put(32,17){\makebox(0,0)[b]{\footnotesize 1-loop}}
\put(32,13){\makebox(0,0)[t]{\mbox{\footnotesize 1PI}}}
\epi
&=&
\bpi(44,11)(-11,-3)
\put(-5,-11){\line(1,0){32}}
\put(11,-11){\circle*{3}}
\put(11,1){\circle{22}}
\epi
+
\bpi(44,11)(-11,-3)
\put(-5,-11){\line(1,0){32}}
\put(11,-11){\circle*{3}}
\put(0,0){\line(0,1){1}}
\multiput(0,1)(0.04,0.2){5}{\line(1,0){0.1}}
\multiput(2.5,7.1)(-0.1,-0.15){6}{\line(1,0){0.1}}
\multiput(2.5,7.1)(0.1,0.1){14}{\line(1,0){0.1}}
\multiput(3.9,8.5)(0.15,0.1){6}{\line(1,0){0.1}}
\multiput(10,11)(-0.2,-0.04){5}{\line(1,0){0.1}}
\put(10,11){\line(1,0){2}}
\multiput(12,11)(0.2,-0.04){5}{\line(1,0){0.1}}
\multiput(19.5,7.1)(0.1,-0.15){6}{\line(1,0){0.1}}
\multiput(19.5,7.1)(-0.1,0.1){14}{\line(1,0){0.1}}
\multiput(18.1,8.5)(-0.15,0.1){6}{\line(1,0){0.1}}
\multiput(22,1)(-0.04,0.2){5}{\line(1,0){0.1}}
\put(22,0){\line(0,1){1}}
\put(0,0){\line(0,-1){1}}
\multiput(0,-1)(0.04,-0.2){5}{\line(1,0){0.1}}
\multiput(2.5,-7.1)(-0.1,0.15){6}{\line(1,0){0.1}}
\multiput(2.5,-7.1)(0.1,-0.1){14}{\line(1,0){0.1}}
\multiput(3.9,-8.5)(0.15,-0.1){6}{\line(1,0){0.1}}
\multiput(10,-11)(-0.2,0.04){5}{\line(1,0){0.1}}
\put(10,-11){\line(1,0){2}}
\multiput(12,-11)(0.2,0.04){5}{\line(1,0){0.1}}
\multiput(19.5,-7.1)(0.1,0.15){6}{\line(1,0){0.1}}
\multiput(19.5,-7.1)(-0.1,-0.1){14}{\line(1,0){0.1}}
\multiput(18.1,-8.5)(-0.15,-0.1){6}{\line(1,0){0.1}}
\multiput(22,-1)(-0.04,-0.2){5}{\line(1,0){0.1}}
\put(22,0){\line(0,-1){1}}
\epi
+
\bpi(44,11)(-11,-3)
\put(-5,-11){\line(1,0){32}}
\put(11,-11){\circle*{3}}
\put(2,0){\oval(4,4)[tl]}
\put(2,4){\oval(4,4)[br]}
\multiput(4,4)(-0.1,0.2){5}{\line(1,0){0.1}}
\multiput(3.5,5)(-0.1,0.1){5}{\line(1,0){0.1}}
\multiput(3,5.5)(-0.1,0.2){5}{\line(1,0){0.1}}
\put(4.5,6.5){\oval(4,4)[tl]}
\multiput(4.5,8.5)(0.2,-0.1){5}{\line(1,0){0.1}}
\multiput(5.5,8)(0.1,-0.1){5}{\line(1,0){0.1}}
\multiput(6,7.5)(0.2,-0.1){5}{\line(1,0){0.1}}
\put(7,9){\oval(4,4)[br]}
\put(11,9){\oval(4,4)[t]}
\put(15,9){\oval(4,4)[bl]}
\multiput(15,7)(0.2,0.1){5}{\line(1,0){0.1}}
\multiput(16,7.5)(0.1,0.1){5}{\line(1,0){0.1}}
\multiput(16.5,8)(0.2,0.1){5}{\line(1,0){0.1}}
\put(17.5,6.5){\oval(4,4)[tr]}
\multiput(19.5,6.5)(-0.1,-0.2){5}{\line(1,0){0.1}}
\multiput(19,5.5)(-0.1,-0.1){5}{\line(1,0){0.1}}
\multiput(18.5,5)(-0.1,-0.2){5}{\line(1,0){0.1}}
\put(20,4){\oval(4,4)[bl]}
\put(20,0){\oval(4,4)[tr]}
\put(2,0){\oval(4,4)[bl]}
\put(2,-4){\oval(4,4)[tr]}
\multiput(4,-4)(-0.1,-0.2){5}{\line(1,0){0.1}}
\multiput(3.5,-5)(-0.1,-0.1){5}{\line(1,0){0.1}}
\multiput(3,-5.5)(-0.1,-0.2){5}{\line(1,0){0.1}}
\put(4.5,-6.5){\oval(4,4)[bl]}
\multiput(4.5,-8.5)(0.2,0.1){5}{\line(1,0){0.1}}
\multiput(5.5,-8)(0.1,0.1){5}{\line(1,0){0.1}}
\multiput(6,-7.5)(0.2,0.1){5}{\line(1,0){0.1}}
\put(7,-9){\oval(4,4)[tr]}
\put(11,-9){\oval(4,4)[b]}
\put(15,-9){\oval(4,4)[tl]}
\multiput(15,-7)(0.2,-0.1){5}{\line(1,0){0.1}}
\multiput(16,-7.5)(0.1,-0.1){5}{\line(1,0){0.1}}
\multiput(16.5,-8)(0.2,-0.1){5}{\line(1,0){0.1}}
\put(17.5,-6.5){\oval(4,4)[br]}
\multiput(19.5,-6.5)(-0.1,0.2){5}{\line(1,0){0.1}}
\multiput(19,-5.5)(-0.1,0.1){5}{\line(1,0){0.1}}
\multiput(18.5,-5)(-0.1,0.2){5}{\line(1,0){0.1}}
\put(20,-4){\oval(4,4)[tl]}
\put(20,0){\oval(4,4)[br]}
\epi
+
\bpi(44,11)(-11,-3)
\put(-5,-11){\line(1,0){32}}
\put(11,-11){\circle*{3}}
\put(0,0){\circle*{1}}
\put(0.8,4.2){\circle*{1}}
\put(3.2,7.8){\circle*{1}}
\put(6.8,10.2){\circle*{1}}
\put(11,11){\circle*{1}}
\put(15.2,10.2){\circle*{1}}
\put(18.8,7.8){\circle*{1}}
\put(21.2,4.2){\circle*{1}}
\put(22,0){\circle*{1}}
\multiput(10,11)(0.1,0.1){28}{\line(1,0){0.1}}
\multiput(10,11)(0.1,-0.1){28}{\line(1,0){0.1}}
\put(0,0){\circle*{1}}
\put(0.8,-4.2){\circle*{1}}
\put(3.2,-7.8){\circle*{1}}
\put(6.8,-10.2){\circle*{1}}
\put(11,-11){\circle*{1}}
\put(15.2,-10.2){\circle*{1}}
\put(18.8,-7.8){\circle*{1}}
\put(21.2,-4.2){\circle*{1}}
\put(22,0){\circle*{1}}
\epi
+
\bpi(44,11)(-11,-3)
\put(-11,0){\line(1,0){11}}
\put(22,0){\line(1,0){11}}
\put(0,0){\circle*{3}}
\put(22,0){\circle*{3}}
\put(2,0){\oval(4,4)[tl]}
\put(2,4){\oval(4,4)[br]}
\multiput(4,4)(-0.1,0.2){5}{\line(1,0){0.1}}
\multiput(3.5,5)(-0.1,0.1){5}{\line(1,0){0.1}}
\multiput(3,5.5)(-0.1,0.2){5}{\line(1,0){0.1}}
\put(4.5,6.5){\oval(4,4)[tl]}
\multiput(4.5,8.5)(0.2,-0.1){5}{\line(1,0){0.1}}
\multiput(5.5,8)(0.1,-0.1){5}{\line(1,0){0.1}}
\multiput(6,7.5)(0.2,-0.1){5}{\line(1,0){0.1}}
\put(7,9){\oval(4,4)[br]}
\put(11,9){\oval(4,4)[t]}
\put(15,9){\oval(4,4)[bl]}
\multiput(15,7)(0.2,0.1){5}{\line(1,0){0.1}}
\multiput(16,7.5)(0.1,0.1){5}{\line(1,0){0.1}}
\multiput(16.5,8)(0.2,0.1){5}{\line(1,0){0.1}}
\put(17.5,6.5){\oval(4,4)[tr]}
\multiput(19.5,6.5)(-0.1,-0.2){5}{\line(1,0){0.1}}
\multiput(19,5.5)(-0.1,-0.1){5}{\line(1,0){0.1}}
\multiput(18.5,5)(-0.1,-0.2){5}{\line(1,0){0.1}}
\put(20,4){\oval(4,4)[bl]}
\put(20,0){\oval(4,4)[tr]}
\put(0,0){\line(0,-1){1}}
\multiput(0,-1)(0.04,-0.2){5}{\line(1,0){0.1}}
\multiput(2.5,-7.1)(-0.1,0.15){6}{\line(1,0){0.1}}
\multiput(2.5,-7.1)(0.1,-0.1){14}{\line(1,0){0.1}}
\multiput(3.9,-8.5)(0.15,-0.1){6}{\line(1,0){0.1}}
\multiput(10,-11)(-0.2,0.04){5}{\line(1,0){0.1}}
\put(10,-11){\line(1,0){2}}
\multiput(12,-11)(0.2,0.04){5}{\line(1,0){0.1}}
\multiput(19.5,-7.1)(0.1,0.15){6}{\line(1,0){0.1}}
\multiput(19.5,-7.1)(-0.1,-0.1){14}{\line(1,0){0.1}}
\multiput(18.1,-8.5)(-0.15,-0.1){6}{\line(1,0){0.1}}
\multiput(22,-1)(-0.04,-0.2){5}{\line(1,0){0.1}}
\put(22,0){\line(0,-1){1}}
\epi
\nn
&&
\bpi(-14,35)
\epi
+
\bpi(44,11)(-11,-3)
\put(-11,0){\line(1,0){11}}
\put(22,0){\line(1,0){11}}
\put(-1,0){\circle*{3}}
\put(23,0){\circle*{3}}
\put(11,0){\circle{22}}
\epi
+
\bpi(44,11)(-11,-3)
\put(-11,0){\line(1,0){11}}
\put(22,0){\line(1,0){11}}
\put(0,0){\circle*{3}}
\put(22,0){\circle*{3}}
\put(0,0){\line(0,1){1}}
\multiput(0,1)(0.04,0.2){5}{\line(1,0){0.1}}
\multiput(2.5,7.1)(-0.1,-0.15){6}{\line(1,0){0.1}}
\multiput(2.5,7.1)(0.1,0.1){14}{\line(1,0){0.1}}
\multiput(3.9,8.5)(0.15,0.1){6}{\line(1,0){0.1}}
\multiput(10,11)(-0.2,-0.04){5}{\line(1,0){0.1}}
\put(10,11){\line(1,0){2}}
\multiput(12,11)(0.2,-0.04){5}{\line(1,0){0.1}}
\multiput(19.5,7.1)(0.1,-0.15){6}{\line(1,0){0.1}}
\multiput(19.5,7.1)(-0.1,0.1){14}{\line(1,0){0.1}}
\multiput(18.1,8.5)(-0.15,0.1){6}{\line(1,0){0.1}}
\multiput(22,1)(-0.04,0.2){5}{\line(1,0){0.1}}
\put(22,0){\line(0,1){1}}
\put(0,0){\line(0,-1){1}}
\multiput(0,-1)(0.04,-0.2){5}{\line(1,0){0.1}}
\multiput(2.5,-7.1)(-0.1,0.15){6}{\line(1,0){0.1}}
\multiput(2.5,-7.1)(0.1,-0.1){14}{\line(1,0){0.1}}
\multiput(3.9,-8.5)(0.15,-0.1){6}{\line(1,0){0.1}}
\multiput(10,-11)(-0.2,0.04){5}{\line(1,0){0.1}}
\put(10,-11){\line(1,0){2}}
\multiput(12,-11)(0.2,0.04){5}{\line(1,0){0.1}}
\multiput(19.5,-7.1)(0.1,0.15){6}{\line(1,0){0.1}}
\multiput(19.5,-7.1)(-0.1,-0.1){14}{\line(1,0){0.1}}
\multiput(18.1,-8.5)(-0.15,-0.1){6}{\line(1,0){0.1}}
\multiput(22,-1)(-0.04,-0.2){5}{\line(1,0){0.1}}
\put(22,0){\line(0,-1){1}}
\epi
+
\bpi(44,11)(-11,-3)
\put(-11,0){\line(1,0){11}}
\put(22,0){\line(1,0){11}}
\put(0,0){\circle*{3}}
\put(22,0){\circle*{3}}
\put(2,0){\oval(4,4)[tl]}
\put(2,4){\oval(4,4)[br]}
\multiput(4,4)(-0.1,0.2){5}{\line(1,0){0.1}}
\multiput(3.5,5)(-0.1,0.1){5}{\line(1,0){0.1}}
\multiput(3,5.5)(-0.1,0.2){5}{\line(1,0){0.1}}
\put(4.5,6.5){\oval(4,4)[tl]}
\multiput(4.5,8.5)(0.2,-0.1){5}{\line(1,0){0.1}}
\multiput(5.5,8)(0.1,-0.1){5}{\line(1,0){0.1}}
\multiput(6,7.5)(0.2,-0.1){5}{\line(1,0){0.1}}
\put(7,9){\oval(4,4)[br]}
\put(11,9){\oval(4,4)[t]}
\put(15,9){\oval(4,4)[bl]}
\multiput(15,7)(0.2,0.1){5}{\line(1,0){0.1}}
\multiput(16,7.5)(0.1,0.1){5}{\line(1,0){0.1}}
\multiput(16.5,8)(0.2,0.1){5}{\line(1,0){0.1}}
\put(17.5,6.5){\oval(4,4)[tr]}
\multiput(19.5,6.5)(-0.1,-0.2){5}{\line(1,0){0.1}}
\multiput(19,5.5)(-0.1,-0.1){5}{\line(1,0){0.1}}
\multiput(18.5,5)(-0.1,-0.2){5}{\line(1,0){0.1}}
\put(20,4){\oval(4,4)[bl]}
\put(20,0){\oval(4,4)[tr]}
\put(2,0){\oval(4,4)[bl]}
\put(2,-4){\oval(4,4)[tr]}
\multiput(4,-4)(-0.1,-0.2){5}{\line(1,0){0.1}}
\multiput(3.5,-5)(-0.1,-0.1){5}{\line(1,0){0.1}}
\multiput(3,-5.5)(-0.1,-0.2){5}{\line(1,0){0.1}}
\put(4.5,-6.5){\oval(4,4)[bl]}
\multiput(4.5,-8.5)(0.2,0.1){5}{\line(1,0){0.1}}
\multiput(5.5,-8)(0.1,0.1){5}{\line(1,0){0.1}}
\multiput(6,-7.5)(0.2,0.1){5}{\line(1,0){0.1}}
\put(7,-9){\oval(4,4)[tr]}
\put(11,-9){\oval(4,4)[b]}
\put(15,-9){\oval(4,4)[tl]}
\multiput(15,-7)(0.2,-0.1){5}{\line(1,0){0.1}}
\multiput(16,-7.5)(0.1,-0.1){5}{\line(1,0){0.1}}
\multiput(16.5,-8)(0.2,-0.1){5}{\line(1,0){0.1}}
\put(17.5,-6.5){\oval(4,4)[br]}
\multiput(19.5,-6.5)(-0.1,0.2){5}{\line(1,0){0.1}}
\multiput(19,-5.5)(-0.1,0.1){5}{\line(1,0){0.1}}
\multiput(18.5,-5)(-0.1,0.2){5}{\line(1,0){0.1}}
\put(20,-4){\oval(4,4)[tl]}
\put(20,0){\oval(4,4)[br]}
\epi
+
\bpi(44,11)(-11,-3)
\put(-11,0){\line(1,0){11}}
\put(22,0){\line(1,0){11}}
\put(0,0){\circle*{3}}
\put(22,0){\circle*{3}}
\put(0,0){\circle*{1}}
\put(0.8,4.2){\circle*{1}}
\put(3.2,7.8){\circle*{1}}
\put(6.8,10.2){\circle*{1}}
\put(11,11){\circle*{1}}
\put(15.2,10.2){\circle*{1}}
\put(18.8,7.8){\circle*{1}}
\put(21.2,4.2){\circle*{1}}
\put(22,0){\circle*{1}}
\multiput(10,11)(0.1,0.1){28}{\line(1,0){0.1}}
\multiput(10,11)(0.1,-0.1){28}{\line(1,0){0.1}}
\put(0,0){\circle*{1}}
\put(0.8,-4.2){\circle*{1}}
\put(3.2,-7.8){\circle*{1}}
\put(6.8,-10.2){\circle*{1}}
\put(11,-11){\circle*{1}}
\put(15.2,-10.2){\circle*{1}}
\put(18.8,-7.8){\circle*{1}}
\put(21.2,-4.2){\circle*{1}}
\put(22,0){\circle*{1}}
\multiput(12,-11)(-0.1,0.1){28}{\line(1,0){0.1}}
\multiput(12,-11)(-0.1,-0.1){28}{\line(1,0){0.1}}
\epi
+
\bpi(50,11)(-11,-3)
\put(-11,0){\line(1,0){44}}
\put(11,0){\circle*{3}}
\put(11,0){\circle{5}}
\epi,
\\
\bpi(40,26)(12,17.5)
\put(0,0){\line(1,0){32}}
\put(16,0){\line(0,1){8}}
\put(16,24){\circle{32}}
\put(16,0){\circle*{3}}
\put(16,8){\circle*{3}}
\put(16,25){\makebox(0,0)[b]{\footnotesize 1-loop}}
\put(16,21){\makebox(0,0)[t]{\mbox{\footnotesize 1PI}}}
\epi
&=&
\bpi(0,40)
\epi
\bpi(44,11)(-11,-5.5)
\put(0,-19){\line(1,0){22}}
\put(11,-19){\line(0,1){8}}
\put(11,-19){\circle*{3}}
\put(11,-11){\circle*{3}}
\put(11,1){\circle{22}}
\epi
+
\bpi(44,11)(-11,-5.5)
\put(0,-19){\line(1,0){22}}
\put(11,-19){\line(0,1){8}}
\put(11,-19){\circle*{3}}
\put(11,-11){\circle*{3}}
\put(0,0){\line(0,1){1}}
\multiput(0,1)(0.04,0.2){5}{\line(1,0){0.1}}
\multiput(2.5,7.1)(-0.1,-0.15){6}{\line(1,0){0.1}}
\multiput(2.5,7.1)(0.1,0.1){14}{\line(1,0){0.1}}
\multiput(3.9,8.5)(0.15,0.1){6}{\line(1,0){0.1}}
\multiput(10,11)(-0.2,-0.04){5}{\line(1,0){0.1}}
\put(10,11){\line(1,0){2}}
\multiput(12,11)(0.2,-0.04){5}{\line(1,0){0.1}}
\multiput(19.5,7.1)(0.1,-0.15){6}{\line(1,0){0.1}}
\multiput(19.5,7.1)(-0.1,0.1){14}{\line(1,0){0.1}}
\multiput(18.1,8.5)(-0.15,0.1){6}{\line(1,0){0.1}}
\multiput(22,1)(-0.04,0.2){5}{\line(1,0){0.1}}
\put(22,0){\line(0,1){1}}
\put(0,0){\line(0,-1){1}}
\multiput(0,-1)(0.04,-0.2){5}{\line(1,0){0.1}}
\multiput(2.5,-7.1)(-0.1,0.15){6}{\line(1,0){0.1}}
\multiput(2.5,-7.1)(0.1,-0.1){14}{\line(1,0){0.1}}
\multiput(3.9,-8.5)(0.15,-0.1){6}{\line(1,0){0.1}}
\multiput(10,-11)(-0.2,0.04){5}{\line(1,0){0.1}}
\put(10,-11){\line(1,0){2}}
\multiput(12,-11)(0.2,0.04){5}{\line(1,0){0.1}}
\multiput(19.5,-7.1)(0.1,0.15){6}{\line(1,0){0.1}}
\multiput(19.5,-7.1)(-0.1,-0.1){14}{\line(1,0){0.1}}
\multiput(18.1,-8.5)(-0.15,-0.1){6}{\line(1,0){0.1}}
\multiput(22,-1)(-0.04,-0.2){5}{\line(1,0){0.1}}
\put(22,0){\line(0,-1){1}}
\epi
+
\bpi(44,11)(-11,-5.5)
\put(0,-19){\line(1,0){22}}
\put(11,-19){\line(0,1){8}}
\put(11,-19){\circle*{3}}
\put(11,-11){\circle*{3}}
\put(2,0){\oval(4,4)[tl]}
\put(2,4){\oval(4,4)[br]}
\multiput(4,4)(-0.1,0.2){5}{\line(1,0){0.1}}
\multiput(3.5,5)(-0.1,0.1){5}{\line(1,0){0.1}}
\multiput(3,5.5)(-0.1,0.2){5}{\line(1,0){0.1}}
\put(4.5,6.5){\oval(4,4)[tl]}
\multiput(4.5,8.5)(0.2,-0.1){5}{\line(1,0){0.1}}
\multiput(5.5,8)(0.1,-0.1){5}{\line(1,0){0.1}}
\multiput(6,7.5)(0.2,-0.1){5}{\line(1,0){0.1}}
\put(7,9){\oval(4,4)[br]}
\put(11,9){\oval(4,4)[t]}
\put(15,9){\oval(4,4)[bl]}
\multiput(15,7)(0.2,0.1){5}{\line(1,0){0.1}}
\multiput(16,7.5)(0.1,0.1){5}{\line(1,0){0.1}}
\multiput(16.5,8)(0.2,0.1){5}{\line(1,0){0.1}}
\put(17.5,6.5){\oval(4,4)[tr]}
\multiput(19.5,6.5)(-0.1,-0.2){5}{\line(1,0){0.1}}
\multiput(19,5.5)(-0.1,-0.1){5}{\line(1,0){0.1}}
\multiput(18.5,5)(-0.1,-0.2){5}{\line(1,0){0.1}}
\put(20,4){\oval(4,4)[bl]}
\put(20,0){\oval(4,4)[tr]}
\put(2,0){\oval(4,4)[bl]}
\put(2,-4){\oval(4,4)[tr]}
\multiput(4,-4)(-0.1,-0.2){5}{\line(1,0){0.1}}
\multiput(3.5,-5)(-0.1,-0.1){5}{\line(1,0){0.1}}
\multiput(3,-5.5)(-0.1,-0.2){5}{\line(1,0){0.1}}
\put(4.5,-6.5){\oval(4,4)[bl]}
\multiput(4.5,-8.5)(0.2,0.1){5}{\line(1,0){0.1}}
\multiput(5.5,-8)(0.1,0.1){5}{\line(1,0){0.1}}
\multiput(6,-7.5)(0.2,0.1){5}{\line(1,0){0.1}}
\put(7,-9){\oval(4,4)[tr]}
\put(11,-9){\oval(4,4)[b]}
\put(15,-9){\oval(4,4)[tl]}
\multiput(15,-7)(0.2,-0.1){5}{\line(1,0){0.1}}
\multiput(16,-7.5)(0.1,-0.1){5}{\line(1,0){0.1}}
\multiput(16.5,-8)(0.2,-0.1){5}{\line(1,0){0.1}}
\put(17.5,-6.5){\oval(4,4)[br]}
\multiput(19.5,-6.5)(-0.1,0.2){5}{\line(1,0){0.1}}
\multiput(19,-5.5)(-0.1,0.1){5}{\line(1,0){0.1}}
\multiput(18.5,-5)(-0.1,0.2){5}{\line(1,0){0.1}}
\put(20,-4){\oval(4,4)[tl]}
\put(20,0){\oval(4,4)[br]}
\epi
+
\bpi(44,11)(-11,-5.5)
\put(0,-19){\line(1,0){22}}
\put(11,-19){\line(0,1){8}}
\put(11,-19){\circle*{3}}
\put(11,-11){\circle*{3}}
\put(0,0){\circle*{1}}
\put(0.8,4.2){\circle*{1}}
\put(3.2,7.8){\circle*{1}}
\put(6.8,10.2){\circle*{1}}
\put(11,11){\circle*{1}}
\put(15.2,10.2){\circle*{1}}
\put(18.8,7.8){\circle*{1}}
\put(21.2,4.2){\circle*{1}}
\put(22,0){\circle*{1}}
\multiput(10,11)(0.1,0.1){28}{\line(1,0){0.1}}
\multiput(10,11)(0.1,-0.1){28}{\line(1,0){0.1}}
\put(0,0){\circle*{1}}
\put(0.8,-4.2){\circle*{1}}
\put(3.2,-7.8){\circle*{1}}
\put(6.8,-10.2){\circle*{1}}
\put(11,-11){\circle*{1}}
\put(15.2,-10.2){\circle*{1}}
\put(18.8,-7.8){\circle*{1}}
\put(21.2,-4.2){\circle*{1}}
\put(22,0){\circle*{1}}
\epi
+
\bpi(50,11)(-11,-5.5)
\put(0,-19){\line(1,0){22}}
\put(11,-19){\line(0,1){19}}
\put(11,-19){\circle*{3}}
\put(11,0){\circle*{3}}
\put(11,0){\circle{5}}
\epi,
\\\nonumber
\eea
and where the external legs have to be truncated as for all other graphs
in this paper. Using the Feynman rules given in tables \ref{feynmanrules}
and \ref{counterterms1} one gets after evaluation of the momentum space
integrals (let $\mu$ be the renormalization scale and define $\bar{\mu}$
by $\ln\bar{\mu}=\ln(4\pi\mu)-\ga_{_E}$, where $\ga_{_E}$ is the
Euler-Mascheroni constant)
\bea
\lefteqn{m_{H_1}^2=}\nn
&&-m^2\left[
\left((6\sqrt{3}\pi-28)\la+10g^2-6\frac{g^4}{\la}\right)
+\left(8\la-6g^2\right)\ln\frac{-2m^2}{\bar{\mu}^2}\right.
\nn
&&\left.\mbox{}+\left(2\la-6g^2\right)\ln\frac{g^2}{2\la}
+4\left(\la-2g^2+3\frac{g^4}{\la}\right)
{\textstyle\sqrt{\frac{2g^2}{\la}-1}}
\,\arctan\frac{1}{\sqrt{\frac{2g^2}{\la}-1}}\right],\;\;\;\;
\eea
which as physical quantity is gauge independent as expected
\cite{ApCa..,Ni,AiFr}.

The determination of the physical gauge boson mass proceeds in close
analogy: Its tree-level value is given by
$m_{A_0}^2=g^2v_0^2=-m^2 g^2/\la$. If we write
\beq
-i
\left[
\bpi(90,26)(-13,12)
\put(-9,13){$\mu$}
\put(66,13){$\nu$}
\multiput(2,16)(8,0){2}{\oval(4,4)[t]}
\multiput(6,16)(8,0){2}{\oval(4,4)[b]}
\multiput(50,16)(8,0){2}{\oval(4,4)[t]}
\multiput(54,16)(8,0){2}{\oval(4,4)[b]}
\put(32,16){\circle{32}}
\put(16,16){\circle*{3}}
\put(48,16){\circle*{3}}
\put(32,17){\makebox(0,0)[b]{\footnotesize 1-loop}}
\put(32,13){\makebox(0,0)[t]{\mbox{\footnotesize 1PI}}}
\epi
+
\bpi(66,26)(-17,16.5)
\put(-9,-3){$\mu$}
\put(34,-3){$\nu$}
\multiput(2,0)(8,0){4}{\oval(4,4)[t]}
\multiput(6,0)(8,0){4}{\oval(4,4)[b]}
\put(16,0){\line(0,1){8}}
\put(16,24){\circle{32}}
\put(16,0){\circle*{3}}
\put(16,8){\circle*{3}}
\put(16,25){\makebox(0,0)[b]{\footnotesize 1-loop}}
\put(16,21){\makebox(0,0)[t]{\mbox{\footnotesize 1PI}}}
\epi
\right]
=
\left(g_{\mu\nu}-\frac{k_\mu k_\nu}{k^2}\right)A
+\frac{k_\mu k_\nu}{k^2}B,
\eeq
where in \rxb gauge
\bea
\bpi(40,26)(28,12)
\multiput(2,16)(8,0){2}{\oval(4,4)[t]}
\multiput(6,16)(8,0){2}{\oval(4,4)[b]}
\multiput(50,16)(8,0){2}{\oval(4,4)[t]}
\multiput(54,16)(8,0){2}{\oval(4,4)[b]}
\put(32,16){\circle{32}}
\put(16,16){\circle*{3}}
\put(48,16){\circle*{3}}
\put(32,17){\makebox(0,0)[b]{\footnotesize 1-loop}}
\put(32,13){\makebox(0,0)[t]{\mbox{\footnotesize 1PI}}}
\epi
&=&
\bpi(40,11)(-9,-5)
\multiput(-5,-13)(8,0){5}{\oval(4,4)[t]}
\multiput(-1,-13)(8,0){4}{\oval(4,4)[b]}
\put(11,-11){\circle*{3}}
\put(11,1){\circle{22}}
\epi
+
\bpi(40,11)(-9,-5)
\multiput(-5,-13)(8,0){5}{\oval(4,4)[t]}
\multiput(-1,-13)(8,0){4}{\oval(4,4)[b]}
\put(11,-11){\circle*{3}}
\put(0,0){\line(0,1){1}}
\multiput(0,1)(0.04,0.2){5}{\line(1,0){0.1}}
\multiput(2.5,7.1)(-0.1,-0.15){6}{\line(1,0){0.1}}
\multiput(2.5,7.1)(0.1,0.1){14}{\line(1,0){0.1}}
\multiput(3.9,8.5)(0.15,0.1){6}{\line(1,0){0.1}}
\multiput(10,11)(-0.2,-0.04){5}{\line(1,0){0.1}}
\put(10,11){\line(1,0){2}}
\multiput(12,11)(0.2,-0.04){5}{\line(1,0){0.1}}
\multiput(19.5,7.1)(0.1,-0.15){6}{\line(1,0){0.1}}
\multiput(19.5,7.1)(-0.1,0.1){14}{\line(1,0){0.1}}
\multiput(18.1,8.5)(-0.15,0.1){6}{\line(1,0){0.1}}
\multiput(22,1)(-0.04,0.2){5}{\line(1,0){0.1}}
\put(22,0){\line(0,1){1}}
\put(0,0){\line(0,-1){1}}
\multiput(0,-1)(0.04,-0.2){5}{\line(1,0){0.1}}
\multiput(2.5,-7.1)(-0.1,0.15){6}{\line(1,0){0.1}}
\multiput(2.5,-7.1)(0.1,-0.1){14}{\line(1,0){0.1}}
\multiput(3.9,-8.5)(0.15,-0.1){6}{\line(1,0){0.1}}
\multiput(10,-11)(-0.2,0.04){5}{\line(1,0){0.1}}
\put(10,-11){\line(1,0){2}}
\multiput(12,-11)(0.2,0.04){5}{\line(1,0){0.1}}
\multiput(19.5,-7.1)(0.1,0.15){6}{\line(1,0){0.1}}
\multiput(19.5,-7.1)(-0.1,-0.1){14}{\line(1,0){0.1}}
\multiput(18.1,-8.5)(-0.15,-0.1){6}{\line(1,0){0.1}}
\multiput(22,-1)(-0.04,-0.2){5}{\line(1,0){0.1}}
\put(22,0){\line(0,-1){1}}
\epi
+
\bpi(54,11)(-16,-3)
\multiput(-14,0)(8,0){2}{\oval(4,4)[t]}
\multiput(-10,0)(8,0){2}{\oval(4,4)[b]}
\multiput(24,0)(8,0){2}{\oval(4,4)[t]}
\multiput(28,0)(8,0){2}{\oval(4,4)[b]}
\put(0,0){\circle*{3}}
\put(22,0){\circle*{3}}
\put(11,0){\oval(22,22)[t]}
\put(2,0){\oval(4,4)[bl]}
\put(2,-4){\oval(4,4)[tr]}
\multiput(4,-4)(-0.1,-0.2){5}{\line(1,0){0.1}}
\multiput(3.5,-5)(-0.1,-0.1){5}{\line(1,0){0.1}}
\multiput(3,-5.5)(-0.1,-0.2){5}{\line(1,0){0.1}}
\put(4.5,-6.5){\oval(4,4)[bl]}
\multiput(4.5,-8.5)(0.2,0.1){5}{\line(1,0){0.1}}
\multiput(5.5,-8)(0.1,0.1){5}{\line(1,0){0.1}}
\multiput(6,-7.5)(0.2,0.1){5}{\line(1,0){0.1}}
\put(7,-9){\oval(4,4)[tr]}
\put(11,-9){\oval(4,4)[b]}
\put(15,-9){\oval(4,4)[tl]}
\multiput(15,-7)(0.2,-0.1){5}{\line(1,0){0.1}}
\multiput(16,-7.5)(0.1,-0.1){5}{\line(1,0){0.1}}
\multiput(16.5,-8)(0.2,-0.1){5}{\line(1,0){0.1}}
\put(17.5,-6.5){\oval(4,4)[br]}
\multiput(19.5,-6.5)(-0.1,0.2){5}{\line(1,0){0.1}}
\multiput(19,-5.5)(-0.1,0.1){5}{\line(1,0){0.1}}
\multiput(18.5,-5)(-0.1,0.2){5}{\line(1,0){0.1}}
\put(20,-4){\oval(4,4)[tl]}
\put(20,0){\oval(4,4)[br]}
\epi
+
\bpi(54,11)(-16,-3)
\multiput(-14,0)(8,0){2}{\oval(4,4)[t]}
\multiput(-10,0)(8,0){2}{\oval(4,4)[b]}
\multiput(24,0)(8,0){2}{\oval(4,4)[t]}
\multiput(28,0)(8,0){2}{\oval(4,4)[b]}
\put(0,0){\circle*{3}}
\put(22,0){\circle*{3}}
\put(11,0){\oval(22,22)[t]}
\put(0,0){\line(0,-1){1}}
\multiput(0,-1)(0.04,-0.2){5}{\line(1,0){0.1}}
\multiput(2.5,-7.1)(-0.1,0.15){6}{\line(1,0){0.1}}
\multiput(2.5,-7.1)(0.1,-0.1){14}{\line(1,0){0.1}}
\multiput(3.9,-8.5)(0.15,-0.1){6}{\line(1,0){0.1}}
\multiput(10,-11)(-0.2,0.04){5}{\line(1,0){0.1}}
\put(10,-11){\line(1,0){2}}
\multiput(12,-11)(0.2,0.04){5}{\line(1,0){0.1}}
\multiput(19.5,-7.1)(0.1,0.15){6}{\line(1,0){0.1}}
\multiput(19.5,-7.1)(-0.1,-0.1){14}{\line(1,0){0.1}}
\multiput(18.1,-8.5)(-0.15,-0.1){6}{\line(1,0){0.1}}
\multiput(22,-1)(-0.04,-0.2){5}{\line(1,0){0.1}}
\put(22,0){\line(0,-1){1}}
\epi
+
\bpi(37,0)
\multiput(2,3)(8,0){4}{\oval(4,4)[t]}
\multiput(6,3)(8,0){4}{\oval(4,4)[b]}
\put(16,3){\circle*{3}}
\put(16,3){\circle{5}}
\epi
,
\\
\bpi(40,26)(12,16.5)
\multiput(2,0)(8,0){4}{\oval(4,4)[t]}
\multiput(6,0)(8,0){4}{\oval(4,4)[b]}
\put(16,0){\line(0,1){8}}
\put(16,24){\circle{32}}
\put(16,0){\circle*{3}}
\put(16,8){\circle*{3}}
\put(16,25){\makebox(0,0)[b]{\footnotesize 1-loop}}
\put(16,21){\makebox(0,0)[t]{\mbox{\footnotesize 1PI}}}
\epi
&=&
\bpi(0,40)
\epi
\bpi(40,11)(-9,-5.5)
\multiput(-3,-19)(8,0){4}{\oval(4,4)[t]}
\multiput(1,-19)(8,0){4}{\oval(4,4)[b]}
\put(11,-19){\line(0,1){8}}
\put(11,-19){\circle*{3}}
\put(11,-11){\circle*{3}}
\put(11,1){\circle{22}}
\epi
+
\bpi(40,11)(-9,-5.5)
\multiput(-3,-19)(8,0){4}{\oval(4,4)[t]}
\multiput(1,-19)(8,0){4}{\oval(4,4)[b]}
\put(11,-19){\line(0,1){8}}
\put(11,-19){\circle*{3}}
\put(11,-11){\circle*{3}}
\put(0,0){\line(0,1){1}}
\multiput(0,1)(0.04,0.2){5}{\line(1,0){0.1}}
\multiput(2.5,7.1)(-0.1,-0.15){6}{\line(1,0){0.1}}
\multiput(2.5,7.1)(0.1,0.1){14}{\line(1,0){0.1}}
\multiput(3.9,8.5)(0.15,0.1){6}{\line(1,0){0.1}}
\multiput(10,11)(-0.2,-0.04){5}{\line(1,0){0.1}}
\put(10,11){\line(1,0){2}}
\multiput(12,11)(0.2,-0.04){5}{\line(1,0){0.1}}
\multiput(19.5,7.1)(0.1,-0.15){6}{\line(1,0){0.1}}
\multiput(19.5,7.1)(-0.1,0.1){14}{\line(1,0){0.1}}
\multiput(18.1,8.5)(-0.15,0.1){6}{\line(1,0){0.1}}
\multiput(22,1)(-0.04,0.2){5}{\line(1,0){0.1}}
\put(22,0){\line(0,1){1}}
\put(0,0){\line(0,-1){1}}
\multiput(0,-1)(0.04,-0.2){5}{\line(1,0){0.1}}
\multiput(2.5,-7.1)(-0.1,0.15){6}{\line(1,0){0.1}}
\multiput(2.5,-7.1)(0.1,-0.1){14}{\line(1,0){0.1}}
\multiput(3.9,-8.5)(0.15,-0.1){6}{\line(1,0){0.1}}
\multiput(10,-11)(-0.2,0.04){5}{\line(1,0){0.1}}
\put(10,-11){\line(1,0){2}}
\multiput(12,-11)(0.2,0.04){5}{\line(1,0){0.1}}
\multiput(19.5,-7.1)(0.1,0.15){6}{\line(1,0){0.1}}
\multiput(19.5,-7.1)(-0.1,-0.1){14}{\line(1,0){0.1}}
\multiput(18.1,-8.5)(-0.15,-0.1){6}{\line(1,0){0.1}}
\multiput(22,-1)(-0.04,-0.2){5}{\line(1,0){0.1}}
\put(22,0){\line(0,-1){1}}
\epi
+
\bpi(40,11)(-9,-5.5)
\multiput(-3,-19)(8,0){4}{\oval(4,4)[t]}
\multiput(1,-19)(8,0){4}{\oval(4,4)[b]}
\put(11,-19){\line(0,1){8}}
\put(11,-19){\circle*{3}}
\put(11,-11){\circle*{3}}
\put(2,0){\oval(4,4)[tl]}
\put(2,4){\oval(4,4)[br]}
\multiput(4,4)(-0.1,0.2){5}{\line(1,0){0.1}}
\multiput(3.5,5)(-0.1,0.1){5}{\line(1,0){0.1}}
\multiput(3,5.5)(-0.1,0.2){5}{\line(1,0){0.1}}
\put(4.5,6.5){\oval(4,4)[tl]}
\multiput(4.5,8.5)(0.2,-0.1){5}{\line(1,0){0.1}}
\multiput(5.5,8)(0.1,-0.1){5}{\line(1,0){0.1}}
\multiput(6,7.5)(0.2,-0.1){5}{\line(1,0){0.1}}
\put(7,9){\oval(4,4)[br]}
\put(11,9){\oval(4,4)[t]}
\put(15,9){\oval(4,4)[bl]}
\multiput(15,7)(0.2,0.1){5}{\line(1,0){0.1}}
\multiput(16,7.5)(0.1,0.1){5}{\line(1,0){0.1}}
\multiput(16.5,8)(0.2,0.1){5}{\line(1,0){0.1}}
\put(17.5,6.5){\oval(4,4)[tr]}
\multiput(19.5,6.5)(-0.1,-0.2){5}{\line(1,0){0.1}}
\multiput(19,5.5)(-0.1,-0.1){5}{\line(1,0){0.1}}
\multiput(18.5,5)(-0.1,-0.2){5}{\line(1,0){0.1}}
\put(20,4){\oval(4,4)[bl]}
\put(20,0){\oval(4,4)[tr]}
\put(2,0){\oval(4,4)[bl]}
\put(2,-4){\oval(4,4)[tr]}
\multiput(4,-4)(-0.1,-0.2){5}{\line(1,0){0.1}}
\multiput(3.5,-5)(-0.1,-0.1){5}{\line(1,0){0.1}}
\multiput(3,-5.5)(-0.1,-0.2){5}{\line(1,0){0.1}}
\put(4.5,-6.5){\oval(4,4)[bl]}
\multiput(4.5,-8.5)(0.2,0.1){5}{\line(1,0){0.1}}
\multiput(5.5,-8)(0.1,0.1){5}{\line(1,0){0.1}}
\multiput(6,-7.5)(0.2,0.1){5}{\line(1,0){0.1}}
\put(7,-9){\oval(4,4)[tr]}
\put(11,-9){\oval(4,4)[b]}
\put(15,-9){\oval(4,4)[tl]}
\multiput(15,-7)(0.2,-0.1){5}{\line(1,0){0.1}}
\multiput(16,-7.5)(0.1,-0.1){5}{\line(1,0){0.1}}
\multiput(16.5,-8)(0.2,-0.1){5}{\line(1,0){0.1}}
\put(17.5,-6.5){\oval(4,4)[br]}
\multiput(19.5,-6.5)(-0.1,0.2){5}{\line(1,0){0.1}}
\multiput(19,-5.5)(-0.1,0.1){5}{\line(1,0){0.1}}
\multiput(18.5,-5)(-0.1,0.2){5}{\line(1,0){0.1}}
\put(20,-4){\oval(4,4)[tl]}
\put(20,0){\oval(4,4)[br]}
\epi
+
\bpi(40,11)(-9,-5.5)
\multiput(-3,-19)(8,0){4}{\oval(4,4)[t]}
\multiput(1,-19)(8,0){4}{\oval(4,4)[b]}
\put(11,-19){\line(0,1){8}}
\put(11,-19){\circle*{3}}
\put(11,-11){\circle*{3}}
\put(0,0){\circle*{1}}
\put(0.8,4.2){\circle*{1}}
\put(3.2,7.8){\circle*{1}}
\put(6.8,10.2){\circle*{1}}
\put(11,11){\circle*{1}}
\put(15.2,10.2){\circle*{1}}
\put(18.8,7.8){\circle*{1}}
\put(21.2,4.2){\circle*{1}}
\put(22,0){\circle*{1}}
\multiput(10,11)(0.1,0.1){28}{\line(1,0){0.1}}
\multiput(10,11)(0.1,-0.1){28}{\line(1,0){0.1}}
\put(0,0){\circle*{1}}
\put(0.8,-4.2){\circle*{1}}
\put(3.2,-7.8){\circle*{1}}
\put(6.8,-10.2){\circle*{1}}
\put(11,-11){\circle*{1}}
\put(15.2,-10.2){\circle*{1}}
\put(18.8,-7.8){\circle*{1}}
\put(21.2,-4.2){\circle*{1}}
\put(22,0){\circle*{1}}
\epi
+
\bpi(37,11)(-5,-5.5)
\multiput(-3,-19)(8,0){4}{\oval(4,4)[t]}
\multiput(1,-19)(8,0){4}{\oval(4,4)[b]}
\put(11,-19){\line(0,1){19}}
\put(11,-19){\circle*{3}}
\put(11,0){\circle*{3}}
\put(11,0){\circle{5}}
\epi,
\\\nonumber
\eea
and truncate the external legs, then the one-loop correction to the
physical gauge boson mass is easily seen to be given by
\beq
\left.m_{A_1}^2=A\right|_{\scriptstyle k^2=m_{A_0}^2,\;\vp^2=v_0^2}\;.
\eeq
Using again the rules in tables \ref{feynmanrules} and \ref{counterterms1},
one gets
\bea
\lefteqn{m_{A_1}^2=}\nn
&&\!\!
-m^2\left[\left(-\frac{4}{3}\la+10g^2-\frac{62}{9}\frac{g^4}{\la}
+\frac{g^6}{\la^2}\right)
-\left(6g^2-\frac{10}{3}\frac{g^4}{\la}+3\frac{g^6}{\la^2}\right)
\ln\frac{-g^2m^2}{\la\bar{\mu}^2}\right.\nn
&&\!\!
\left.-\left(\frac{4}{3}\frac{\la^2}{g^2}-4\la\right)
\ln\frac{g^2}{2\la}
+\left(\frac{8}{3}\frac{\la^2}{g^2}-\frac{16}{3}\la+8g^2\right)
{\textstyle\sqrt{\frac{2g^2}{\la}-1}\,
\arctan\sqrt{\frac{2g^2}{\la}-1}}\right],\;\;\;\;\;\;
\eea
which also is gauge independent.

Now we proceed to compute the EP at the one-loop level in both
$R_\xi$ and \rxb gauges. The tree-level potential is
just $V_0=\quarter\la\vp^4+\half m^2\vp^2$. In $R_\xi$
gauge I have computed its one-loop correction by summing up graphs
with all numbers of external lines (i.e.\ using (\ref{epomega}) at
$\om=0$) as well as from vacuum graphs as in \cite{Ja,AiFr,DoJa}, but as
stated earlier the price one pays in the latter case is the use of
mixed propagators between longitudinal gauge boson modes and
would-be-Goldstone bosons. The result is of course the same:
\bea
\label{v1rxi}
V_{1,R_\xi}(\vp)
&=&
\frac{1}{4(4\pi)^2}\left[m_H^4\left(\ln\frac{m_H^2}{\bar{\mu}^2}-\frac{3}{2}
\right)+m_A^4\left(3\ln\frac{m_A^2}{\bar{\mu}^2}-\frac{5}{2}\right)\right.\nn
&&\left.\mbox{}+m_a^4\left(\ln\frac{m_a^2}{\bar{\mu}^2}-\frac{3}{2}\right)
+m_b^4\left(\ln\frac{m_b^2}{\bar{\mu}^2}-\frac{3}{2}\right)
-2m_c^4\left(\ln\frac{m_c^2}{\bar{\mu}^2}-\frac{3}{2}\right)\right]\;\;\;\;\;\;
\eea
with (the tree-level values of the squares of) the Higgs mass
$m_H^2=3\la\vp^2+m^2$ and the physical gauge boson mass
$m_A^2=g^2\vp^2$ and with
$m_{a,b}^2=\half(\la\vp^2+m^2)+\xi g^2\vp v_0$
$\pm\half\sqrt{(\la\vp^2+m^2)[(\la\vp^2+m^2)
-4\xi g^2\vp(\vp-v_0)]}$ and $m_c^2=\xi g^2\vp v_0$.

In \rxb gauge, we simply have
\bea
\label{v1rxibar}
V_{1,\overline{R}_\xi}(\vp)
&=&
i
\bpi(44,20)(0,16.5)
\put(22,20){\circle{32}}
\put(22,21){\makebox(0,0)[b]{\footnotesize 1-loop}}
\put(22,17){\makebox(0,0)[t]{\mbox{\footnotesize 1PI}}}
\epi
\nn
\bpi(0,30)
\epi
&=&
i\left[
\bpi(44,11)(-11,-3)
\put(11,1){\circle{22}}
\epi
+
\bpi(44,11)(-11,-3)
\put(0,0){\line(0,1){1}}
\multiput(0,1)(0.04,0.2){5}{\line(1,0){0.1}}
\multiput(2.5,7.1)(-0.1,-0.15){6}{\line(1,0){0.1}}
\multiput(2.5,7.1)(0.1,0.1){14}{\line(1,0){0.1}}
\multiput(3.9,8.5)(0.15,0.1){6}{\line(1,0){0.1}}
\multiput(10,11)(-0.2,-0.04){5}{\line(1,0){0.1}}
\put(10,11){\line(1,0){2}}
\multiput(12,11)(0.2,-0.04){5}{\line(1,0){0.1}}
\multiput(19.5,7.1)(0.1,-0.15){6}{\line(1,0){0.1}}
\multiput(19.5,7.1)(-0.1,0.1){14}{\line(1,0){0.1}}
\multiput(18.1,8.5)(-0.15,0.1){6}{\line(1,0){0.1}}
\multiput(22,1)(-0.04,0.2){5}{\line(1,0){0.1}}
\put(22,0){\line(0,1){1}}
\put(0,0){\line(0,-1){1}}
\multiput(0,-1)(0.04,-0.2){5}{\line(1,0){0.1}}
\multiput(2.5,-7.1)(-0.1,0.15){6}{\line(1,0){0.1}}
\multiput(2.5,-7.1)(0.1,-0.1){14}{\line(1,0){0.1}}
\multiput(3.9,-8.5)(0.15,-0.1){6}{\line(1,0){0.1}}
\multiput(10,-11)(-0.2,0.04){5}{\line(1,0){0.1}}
\put(10,-11){\line(1,0){2}}
\multiput(12,-11)(0.2,0.04){5}{\line(1,0){0.1}}
\multiput(19.5,-7.1)(0.1,0.15){6}{\line(1,0){0.1}}
\multiput(19.5,-7.1)(-0.1,-0.1){14}{\line(1,0){0.1}}
\multiput(18.1,-8.5)(-0.15,-0.1){6}{\line(1,0){0.1}}
\multiput(22,-1)(-0.04,-0.2){5}{\line(1,0){0.1}}
\put(22,0){\line(0,-1){1}}
\epi
+
\bpi(44,11)(-11,-3)
\put(2,0){\oval(4,4)[tl]}
\put(2,4){\oval(4,4)[br]}
\multiput(4,4)(-0.1,0.2){5}{\line(1,0){0.1}}
\multiput(3.5,5)(-0.1,0.1){5}{\line(1,0){0.1}}
\multiput(3,5.5)(-0.1,0.2){5}{\line(1,0){0.1}}
\put(4.5,6.5){\oval(4,4)[tl]}
\multiput(4.5,8.5)(0.2,-0.1){5}{\line(1,0){0.1}}
\multiput(5.5,8)(0.1,-0.1){5}{\line(1,0){0.1}}
\multiput(6,7.5)(0.2,-0.1){5}{\line(1,0){0.1}}
\put(7,9){\oval(4,4)[br]}
\put(11,9){\oval(4,4)[t]}
\put(15,9){\oval(4,4)[bl]}
\multiput(15,7)(0.2,0.1){5}{\line(1,0){0.1}}
\multiput(16,7.5)(0.1,0.1){5}{\line(1,0){0.1}}
\multiput(16.5,8)(0.2,0.1){5}{\line(1,0){0.1}}
\put(17.5,6.5){\oval(4,4)[tr]}
\multiput(19.5,6.5)(-0.1,-0.2){5}{\line(1,0){0.1}}
\multiput(19,5.5)(-0.1,-0.1){5}{\line(1,0){0.1}}
\multiput(18.5,5)(-0.1,-0.2){5}{\line(1,0){0.1}}
\put(20,4){\oval(4,4)[bl]}
\put(20,0){\oval(4,4)[tr]}
\put(2,0){\oval(4,4)[bl]}
\put(2,-4){\oval(4,4)[tr]}
\multiput(4,-4)(-0.1,-0.2){5}{\line(1,0){0.1}}
\multiput(3.5,-5)(-0.1,-0.1){5}{\line(1,0){0.1}}
\multiput(3,-5.5)(-0.1,-0.2){5}{\line(1,0){0.1}}
\put(4.5,-6.5){\oval(4,4)[bl]}
\multiput(4.5,-8.5)(0.2,0.1){5}{\line(1,0){0.1}}
\multiput(5.5,-8)(0.1,0.1){5}{\line(1,0){0.1}}
\multiput(6,-7.5)(0.2,0.1){5}{\line(1,0){0.1}}
\put(7,-9){\oval(4,4)[tr]}
\put(11,-9){\oval(4,4)[b]}
\put(15,-9){\oval(4,4)[tl]}
\multiput(15,-7)(0.2,-0.1){5}{\line(1,0){0.1}}
\multiput(16,-7.5)(0.1,-0.1){5}{\line(1,0){0.1}}
\multiput(16.5,-8)(0.2,-0.1){5}{\line(1,0){0.1}}
\put(17.5,-6.5){\oval(4,4)[br]}
\multiput(19.5,-6.5)(-0.1,0.2){5}{\line(1,0){0.1}}
\multiput(19,-5.5)(-0.1,0.1){5}{\line(1,0){0.1}}
\multiput(18.5,-5)(-0.1,0.2){5}{\line(1,0){0.1}}
\put(20,-4){\oval(4,4)[tl]}
\put(20,0){\oval(4,4)[br]}
\epi
+
\bpi(44,11)(-11,-3)
\put(0,0){\circle*{1}}
\put(0.8,4.2){\circle*{1}}
\put(3.2,7.8){\circle*{1}}
\put(6.8,10.2){\circle*{1}}
\put(11,11){\circle*{1}}
\put(15.2,10.2){\circle*{1}}
\put(18.8,7.8){\circle*{1}}
\put(21.2,4.2){\circle*{1}}
\put(22,0){\circle*{1}}
\multiput(10,11)(0.1,0.1){28}{\line(1,0){0.1}}
\multiput(10,11)(0.1,-0.1){28}{\line(1,0){0.1}}
\put(0,0){\circle*{1}}
\put(0.8,-4.2){\circle*{1}}
\put(3.2,-7.8){\circle*{1}}
\put(6.8,-10.2){\circle*{1}}
\put(11,-11){\circle*{1}}
\put(15.2,-10.2){\circle*{1}}
\put(18.8,-7.8){\circle*{1}}
\put(21.2,-4.2){\circle*{1}}
\put(22,0){\circle*{1}}
\epi
+
\bpi(50,11)(-11,-3)
\put(11,0){\circle*{3}}
\put(11,0){\circle{5}}
\epi
\right]
\nn
&=&\bpi(0,30)\epi\frac{1}{4(4\pi)^2}\left[
m_H^4\left(\ln\frac{m_H^2}{\bar{\mu}^2}-\frac{3}{2}\right)
+m_A^4\left(3\ln\frac{m_A^2}{\bar{\mu}^2}-\frac{5}{2}\right)\right.
\nn
&&
\left.\;\;\;\;\;\;\;\;\;\;\;\;\mbox{}
+m_G^4\left(\ln\frac{m_G^2}{\bar{\mu}^2}-\frac{3}{2}\right)
-m_{\gh}^4\left(\ln\frac{m_{\gh}^2}{\bar{\mu}^2}-\frac{3}{2}\right)
\right]
\eea
with the would-be-Goldstone mass $m_G^2=(\la+\xi g^2)\vp^2+m^2$,
the ghost (and longitudinal gauge boson) mass $m_{\gh}^2=\xi g^2\vp^2$
and $m_H^2$ and $m_A^2$ as above. Clearly, in \rxb gauge $V_1$
is not only easier to compute but also has a more convenient analytical
structure due to the simple expressions for the appearing masses which
contain no awkward square roots anymore.

It is straightforward to determine the two-loop contribution to the EP as
\bea
\label{v2rxibar}
\lefteqn{V_{2,\overline{R}_\xi}(\vp)\;=\;i
\bpi(44,20)(0,16.5)
\put(22,20){\circle{32}}
\put(22,21){\makebox(0,0)[b]{\footnotesize 2-loop}}
\put(22,17){\makebox(0,0)[t]{\mbox{\footnotesize 1PI}}}
\epi
}
\nn
\bpi(0,40)
\epi
&=&
i\left[
\bpi(33,30)(-6,9)
\put(11,1){\circle{22}}
\put(11,25){\circle{22}}
\put(11,13){\circle*{3}}
\epi
+
\bpi(33,11)(-6,9)
\put(11,1){\circle{22}}
\put(0,24){\line(0,1){1}}
\multiput(0,26)(0.04,0.2){5}{\line(1,0){0.1}}
\multiput(2.5,31.1)(-0.1,-0.15){6}{\line(1,0){0.1}}
\multiput(2.5,31.1)(0.1,0.1){14}{\line(1,0){0.1}}
\multiput(3.9,32.5)(0.15,0.1){6}{\line(1,0){0.1}}
\multiput(10,35)(-0.2,-0.04){5}{\line(1,0){0.1}}
\put(10,35){\line(1,0){2}}
\multiput(12,35)(0.2,-0.04){5}{\line(1,0){0.1}}
\multiput(19.5,31.1)(0.1,-0.15){6}{\line(1,0){0.1}}
\multiput(19.5,31.1)(-0.1,0.1){14}{\line(1,0){0.1}}
\multiput(18.1,32.5)(-0.15,0.1){6}{\line(1,0){0.1}}
\multiput(22,25)(-0.04,0.2){5}{\line(1,0){0.1}}
\put(22,24){\line(0,1){1}}
\put(0,24){\line(0,-1){1}}
\multiput(0,23)(0.04,-0.2){5}{\line(1,0){0.1}}
\multiput(2.5,16.9)(-0.1,0.15){6}{\line(1,0){0.1}}
\multiput(2.5,16,9)(0.1,-0.1){14}{\line(1,0){0.1}}
\multiput(3.9,15.5)(0.15,-0.1){6}{\line(1,0){0.1}}
\multiput(10,13)(-0.2,0.04){5}{\line(1,0){0.1}}
\put(10,13){\line(1,0){2}}
\multiput(12,13)(0.2,0.04){5}{\line(1,0){0.1}}
\multiput(19.5,16.9)(0.1,0.15){6}{\line(1,0){0.1}}
\multiput(19.5,16.9)(-0.1,-0.1){14}{\line(1,0){0.1}}
\multiput(18.1,15.5)(-0.15,-0.1){6}{\line(1,0){0.1}}
\multiput(22,23)(-0.04,-0.2){5}{\line(1,0){0.1}}
\put(22,24){\line(0,-1){1}}
\put(11,13){\circle*{3}}
\epi
+
\bpi(33,11)(-6,9)
\put(0,2){\line(0,1){1}}
\multiput(0,3)(0.04,0.2){5}{\line(1,0){0.1}}
\multiput(2.5,9.1)(-0.1,-0.15){6}{\line(1,0){0.1}}
\multiput(2.5,9.1)(0.1,0.1){14}{\line(1,0){0.1}}
\multiput(3.9,10.5)(0.15,0.1){6}{\line(1,0){0.1}}
\multiput(10,13)(-0.2,-0.04){5}{\line(1,0){0.1}}
\put(10,13){\line(1,0){2}}
\multiput(12,13)(0.2,-0.04){5}{\line(1,0){0.1}}
\multiput(19.5,9.1)(0.1,-0.15){6}{\line(1,0){0.1}}
\multiput(19.5,9.1)(-0.1,0.1){14}{\line(1,0){0.1}}
\multiput(18.1,10.5)(-0.15,0.1){6}{\line(1,0){0.1}}
\multiput(22,3)(-0.04,0.2){5}{\line(1,0){0.1}}
\put(22,2){\line(0,1){1}}
\put(0,2){\line(0,-1){1}}
\multiput(0,1)(0.04,-0.2){5}{\line(1,0){0.1}}
\multiput(2.5,-5.1)(-0.1,0.15){6}{\line(1,0){0.1}}
\multiput(2.5,-5.1)(0.1,-0.1){14}{\line(1,0){0.1}}
\multiput(3.9,-6.5)(0.15,-0.1){6}{\line(1,0){0.1}}
\multiput(10,-9)(-0.2,0.04){5}{\line(1,0){0.1}}
\put(10,-9){\line(1,0){2}}
\multiput(12,-9)(0.2,0.04){5}{\line(1,0){0.1}}
\multiput(19.5,-5.1)(0.1,0.15){6}{\line(1,0){0.1}}
\multiput(19.5,-5.1)(-0.1,-0.1){14}{\line(1,0){0.1}}
\multiput(18.1,-6.5)(-0.15,-0.1){6}{\line(1,0){0.1}}
\multiput(22,1)(-0.04,-0.2){5}{\line(1,0){0.1}}
\put(22,2){\line(0,-1){1}}
\put(0,24){\line(0,1){1}}
\multiput(0,25)(0.04,0.2){5}{\line(1,0){0.1}}
\multiput(2.5,31.1)(-0.1,-0.15){6}{\line(1,0){0.1}}
\multiput(2.5,31.1)(0.1,0.1){14}{\line(1,0){0.1}}
\multiput(3.9,32.5)(0.15,0.1){6}{\line(1,0){0.1}}
\multiput(10,35)(-0.2,-0.04){5}{\line(1,0){0.1}}
\put(10,35){\line(1,0){2}}
\multiput(12,35)(0.2,-0.04){5}{\line(1,0){0.1}}
\multiput(19.5,31.1)(0.1,-0.15){6}{\line(1,0){0.1}}
\multiput(19.5,31.1)(-0.1,0.1){14}{\line(1,0){0.1}}
\multiput(18.1,32.5)(-0.15,0.1){6}{\line(1,0){0.1}}
\multiput(22,25)(-0.04,0.2){5}{\line(1,0){0.1}}
\put(22,24){\line(0,1){1}}
\put(0,24){\line(0,-1){1}}
\multiput(0,23)(0.04,-0.2){5}{\line(1,0){0.1}}
\multiput(2.5,16.9)(-0.1,0.15){6}{\line(1,0){0.1}}
\multiput(2.5,16,9)(0.1,-0.1){14}{\line(1,0){0.1}}
\multiput(3.9,15.5)(0.15,-0.1){6}{\line(1,0){0.1}}
\multiput(10,13)(-0.2,0.04){5}{\line(1,0){0.1}}
\put(10,13){\line(1,0){2}}
\multiput(12,13)(0.2,0.04){5}{\line(1,0){0.1}}
\multiput(19.5,16.9)(0.1,0.15){6}{\line(1,0){0.1}}
\multiput(19.5,16.9)(-0.1,-0.1){14}{\line(1,0){0.1}}
\multiput(18.1,15.5)(-0.15,-0.1){6}{\line(1,0){0.1}}
\multiput(22,23)(-0.04,-0.2){5}{\line(1,0){0.1}}
\put(22,24){\line(0,-1){1}}
\put(11,13){\circle*{3}}
\epi
+
\bpi(33,11)(-6,9)
\put(11,1){\circle{22}}
\put(11,13){\circle*{3}}
\put(2,24){\oval(4,4)[tl]}
\put(2,28){\oval(4,4)[br]}
\multiput(4,28)(-0.1,0.2){5}{\line(1,0){0.1}}
\multiput(3.5,29)(-0.1,0.1){5}{\line(1,0){0.1}}
\multiput(3,29.5)(-0.1,0.2){5}{\line(1,0){0.1}}
\put(4.5,30.5){\oval(4,4)[tl]}
\multiput(4.5,32.5)(0.2,-0.1){5}{\line(1,0){0.1}}
\multiput(5.5,32)(0.1,-0.1){5}{\line(1,0){0.1}}
\multiput(6,31.5)(0.2,-0.1){5}{\line(1,0){0.1}}
\put(7,33){\oval(4,4)[br]}
\put(11,33){\oval(4,4)[t]}
\put(15,33){\oval(4,4)[bl]}
\multiput(15,31)(0.2,0.1){5}{\line(1,0){0.1}}
\multiput(16,31.5)(0.1,0.1){5}{\line(1,0){0.1}}
\multiput(16.5,32)(0.2,0.1){5}{\line(1,0){0.1}}
\put(17.5,30.5){\oval(4,4)[tr]}
\multiput(19.5,30.5)(-0.1,-0.2){5}{\line(1,0){0.1}}
\multiput(19,29.5)(-0.1,-0.1){5}{\line(1,0){0.1}}
\multiput(18.5,29)(-0.1,-0.2){5}{\line(1,0){0.1}}
\put(20,28){\oval(4,4)[bl]}
\put(20,24){\oval(4,4)[tr]}
\put(2,24){\oval(4,4)[bl]}
\put(2,20){\oval(4,4)[tr]}
\multiput(4,20)(-0.1,-0.2){5}{\line(1,0){0.1}}
\multiput(3.5,19)(-0.1,-0.1){5}{\line(1,0){0.1}}
\multiput(3,18.5)(-0.1,-0.2){5}{\line(1,0){0.1}}
\put(4.5,17.5){\oval(4,4)[bl]}
\multiput(4.5,15.5)(0.2,0.1){5}{\line(1,0){0.1}}
\multiput(5.5,16)(0.1,0.1){5}{\line(1,0){0.1}}
\multiput(6,16.5)(0.2,0.1){5}{\line(1,0){0.1}}
\put(7,15){\oval(4,4)[tr]}
\put(11,15){\oval(4,4)[b]}
\put(15,15){\oval(4,4)[tl]}
\multiput(15,17)(0.2,-0.1){5}{\line(1,0){0.1}}
\multiput(16,16.5)(0.1,-0.1){5}{\line(1,0){0.1}}
\multiput(16.5,16)(0.2,-0.1){5}{\line(1,0){0.1}}
\put(17.5,17.5){\oval(4,4)[br]}
\multiput(19.5,17.5)(-0.1,0.2){5}{\line(1,0){0.1}}
\multiput(19,18.5)(-0.1,0.1){5}{\line(1,0){0.1}}
\multiput(18.5,19)(-0.1,0.2){5}{\line(1,0){0.1}}
\put(20,20){\oval(4,4)[tl]}
\put(20,24){\oval(4,4)[br]}
\epi
+
\bpi(33,11)(-6,9)
\put(0,2){\line(0,1){1}}
\multiput(0,3)(0.04,0.2){5}{\line(1,0){0.1}}
\multiput(2.5,9.1)(-0.1,-0.15){6}{\line(1,0){0.1}}
\multiput(2.5,9.1)(0.1,0.1){14}{\line(1,0){0.1}}
\multiput(3.9,10.5)(0.15,0.1){6}{\line(1,0){0.1}}
\multiput(10,13)(-0.2,-0.04){5}{\line(1,0){0.1}}
\put(10,13){\line(1,0){2}}
\multiput(12,13)(0.2,-0.04){5}{\line(1,0){0.1}}
\multiput(19.5,9.1)(0.1,-0.15){6}{\line(1,0){0.1}}
\multiput(19.5,9.1)(-0.1,0.1){14}{\line(1,0){0.1}}
\multiput(18.1,10.5)(-0.15,0.1){6}{\line(1,0){0.1}}
\multiput(22,3)(-0.04,0.2){5}{\line(1,0){0.1}}
\put(22,2){\line(0,1){1}}
\put(0,2){\line(0,-1){1}}
\multiput(0,1)(0.04,-0.2){5}{\line(1,0){0.1}}
\multiput(2.5,-5.1)(-0.1,0.15){6}{\line(1,0){0.1}}
\multiput(2.5,-5.1)(0.1,-0.1){14}{\line(1,0){0.1}}
\multiput(3.9,-6.5)(0.15,-0.1){6}{\line(1,0){0.1}}
\multiput(10,-9)(-0.2,0.04){5}{\line(1,0){0.1}}
\put(10,-9){\line(1,0){2}}
\multiput(12,-9)(0.2,0.04){5}{\line(1,0){0.1}}
\multiput(19.5,-5.1)(0.1,0.15){6}{\line(1,0){0.1}}
\multiput(19.5,-5.1)(-0.1,-0.1){14}{\line(1,0){0.1}}
\multiput(18.1,-6.5)(-0.15,-0.1){6}{\line(1,0){0.1}}
\multiput(22,1)(-0.04,-0.2){5}{\line(1,0){0.1}}
\put(22,2){\line(0,-1){1}}
\put(11,13){\circle*{3}}
\put(2,24){\oval(4,4)[tl]}
\put(2,28){\oval(4,4)[br]}
\multiput(4,28)(-0.1,0.2){5}{\line(1,0){0.1}}
\multiput(3.5,29)(-0.1,0.1){5}{\line(1,0){0.1}}
\multiput(3,29.5)(-0.1,0.2){5}{\line(1,0){0.1}}
\put(4.5,30.5){\oval(4,4)[tl]}
\multiput(4.5,32.5)(0.2,-0.1){5}{\line(1,0){0.1}}
\multiput(5.5,32)(0.1,-0.1){5}{\line(1,0){0.1}}
\multiput(6,31.5)(0.2,-0.1){5}{\line(1,0){0.1}}
\put(7,33){\oval(4,4)[br]}
\put(11,33){\oval(4,4)[t]}
\put(15,33){\oval(4,4)[bl]}
\multiput(15,31)(0.2,0.1){5}{\line(1,0){0.1}}
\multiput(16,31.5)(0.1,0.1){5}{\line(1,0){0.1}}
\multiput(16.5,32)(0.2,0.1){5}{\line(1,0){0.1}}
\put(17.5,30.5){\oval(4,4)[tr]}
\multiput(19.5,30.5)(-0.1,-0.2){5}{\line(1,0){0.1}}
\multiput(19,29.5)(-0.1,-0.1){5}{\line(1,0){0.1}}
\multiput(18.5,29)(-0.1,-0.2){5}{\line(1,0){0.1}}
\put(20,28){\oval(4,4)[bl]}
\put(20,24){\oval(4,4)[tr]}
\put(2,24){\oval(4,4)[bl]}
\put(2,20){\oval(4,4)[tr]}
\multiput(4,20)(-0.1,-0.2){5}{\line(1,0){0.1}}
\multiput(3.5,19)(-0.1,-0.1){5}{\line(1,0){0.1}}
\multiput(3,18.5)(-0.1,-0.2){5}{\line(1,0){0.1}}
\put(4.5,17.5){\oval(4,4)[bl]}
\multiput(4.5,15.5)(0.2,0.1){5}{\line(1,0){0.1}}
\multiput(5.5,16)(0.1,0.1){5}{\line(1,0){0.1}}
\multiput(6,16.5)(0.2,0.1){5}{\line(1,0){0.1}}
\put(7,15){\oval(4,4)[tr]}
\put(11,15){\oval(4,4)[b]}
\put(15,15){\oval(4,4)[tl]}
\multiput(15,17)(0.2,-0.1){5}{\line(1,0){0.1}}
\multiput(16,16.5)(0.1,-0.1){5}{\line(1,0){0.1}}
\multiput(16.5,16)(0.2,-0.1){5}{\line(1,0){0.1}}
\put(17.5,17.5){\oval(4,4)[br]}
\multiput(19.5,17.5)(-0.1,0.2){5}{\line(1,0){0.1}}
\multiput(19,18.5)(-0.1,0.1){5}{\line(1,0){0.1}}
\multiput(18.5,19)(-0.1,0.2){5}{\line(1,0){0.1}}
\put(20,20){\oval(4,4)[tl]}
\put(20,24){\oval(4,4)[br]}
\epi
+
\bpi(33,11)(-6,9)
\put(11,1){\circle{22}}
\put(11,13){\circle*{3}}
\put(0,24){\circle*{1}}
\put(0.8,28.2){\circle*{1}}
\put(3.2,31.8){\circle*{1}}
\put(6.8,34.2){\circle*{1}}
\put(11,35){\circle*{1}}
\put(15.2,34.2){\circle*{1}}
\put(18.8,31.8){\circle*{1}}
\put(21.2,28.2){\circle*{1}}
\put(22,24){\circle*{1}}
\multiput(10,35)(0.1,0.1){28}{\line(1,0){0.1}}
\multiput(10,35)(0.1,-0.1){28}{\line(1,0){0.1}}
\put(0,24){\circle*{1}}
\put(0.8,19.8){\circle*{1}}
\put(3.2,16.2){\circle*{1}}
\put(6.8,13.8){\circle*{1}}
\put(11,13){\circle*{1}}
\put(15.2,13.8){\circle*{1}}
\put(18.8,16.2){\circle*{1}}
\put(21.2,19.8){\circle*{1}}
\put(22,24){\circle*{1}}
\epi
+
\bpi(33,11)(-6,9)
\put(0,2){\line(0,1){1}}
\multiput(0,3)(0.04,0.2){5}{\line(1,0){0.1}}
\multiput(2.5,9.1)(-0.1,-0.15){6}{\line(1,0){0.1}}
\multiput(2.5,9.1)(0.1,0.1){14}{\line(1,0){0.1}}
\multiput(3.9,10.5)(0.15,0.1){6}{\line(1,0){0.1}}
\multiput(10,13)(-0.2,-0.04){5}{\line(1,0){0.1}}
\put(10,13){\line(1,0){2}}
\multiput(12,13)(0.2,-0.04){5}{\line(1,0){0.1}}
\multiput(19.5,9.1)(0.1,-0.15){6}{\line(1,0){0.1}}
\multiput(19.5,9.1)(-0.1,0.1){14}{\line(1,0){0.1}}
\multiput(18.1,10.5)(-0.15,0.1){6}{\line(1,0){0.1}}
\multiput(22,3)(-0.04,0.2){5}{\line(1,0){0.1}}
\put(22,2){\line(0,1){1}}
\put(0,2){\line(0,-1){1}}
\multiput(0,1)(0.04,-0.2){5}{\line(1,0){0.1}}
\multiput(2.5,-5.1)(-0.1,0.15){6}{\line(1,0){0.1}}
\multiput(2.5,-5.1)(0.1,-0.1){14}{\line(1,0){0.1}}
\multiput(3.9,-6.5)(0.15,-0.1){6}{\line(1,0){0.1}}
\multiput(10,-9)(-0.2,0.04){5}{\line(1,0){0.1}}
\put(10,-9){\line(1,0){2}}
\multiput(12,-9)(0.2,0.04){5}{\line(1,0){0.1}}
\multiput(19.5,-5.1)(0.1,0.15){6}{\line(1,0){0.1}}
\multiput(19.5,-5.1)(-0.1,-0.1){14}{\line(1,0){0.1}}
\multiput(18.1,-6.5)(-0.15,-0.1){6}{\line(1,0){0.1}}
\multiput(22,1)(-0.04,-0.2){5}{\line(1,0){0.1}}
\put(22,2){\line(0,-1){1}}
\put(11,13){\circle*{3}}
\put(0,24){\circle*{1}}
\put(0.8,28.2){\circle*{1}}
\put(3.2,31.8){\circle*{1}}
\put(6.8,34.2){\circle*{1}}
\put(11,35){\circle*{1}}
\put(15.2,34.2){\circle*{1}}
\put(18.8,31.8){\circle*{1}}
\put(21.2,28.2){\circle*{1}}
\put(22,24){\circle*{1}}
\multiput(10,35)(0.1,0.1){28}{\line(1,0){0.1}}
\multiput(10,35)(0.1,-0.1){28}{\line(1,0){0.1}}
\put(0,24){\circle*{1}}
\put(0.8,19.8){\circle*{1}}
\put(3.2,16.2){\circle*{1}}
\put(6.8,13.8){\circle*{1}}
\put(11,13){\circle*{1}}
\put(15.2,13.8){\circle*{1}}
\put(18.8,16.2){\circle*{1}}
\put(21.2,19.8){\circle*{1}}
\put(22,24){\circle*{1}}
\epi
\right.
\nn
\bpi(0,20)
\epi
&&
\left.\;\;\;\;
+
\bpi(33,20)(-6,-3)
\put(11,1){\circle{22}}
\put(-1,0){\circle*{3}}
\put(23,0){\circle*{3}}
\put(-2,0){\line(1,0){24}}
\epi
+
\bpi(33,20)(-6,-3)
\put(0,0){\line(0,1){1}}
\multiput(0,1)(0.04,0.2){5}{\line(1,0){0.1}}
\multiput(2.5,7.1)(-0.1,-0.15){6}{\line(1,0){0.1}}
\multiput(2.5,7.1)(0.1,0.1){14}{\line(1,0){0.1}}
\multiput(3.9,8.5)(0.15,0.1){6}{\line(1,0){0.1}}
\multiput(10,11)(-0.2,-0.04){5}{\line(1,0){0.1}}
\put(10,11){\line(1,0){2}}
\multiput(12,11)(0.2,-0.04){5}{\line(1,0){0.1}}
\multiput(19.5,7.1)(0.1,-0.15){6}{\line(1,0){0.1}}
\multiput(19.5,7.1)(-0.1,0.1){14}{\line(1,0){0.1}}
\multiput(18.1,8.5)(-0.15,0.1){6}{\line(1,0){0.1}}
\multiput(22,1)(-0.04,0.2){5}{\line(1,0){0.1}}
\put(22,0){\line(0,1){1}}
\put(0,0){\line(0,-1){1}}
\multiput(0,-1)(0.04,-0.2){5}{\line(1,0){0.1}}
\multiput(2.5,-7.1)(-0.1,0.15){6}{\line(1,0){0.1}}
\multiput(2.5,-7.1)(0.1,-0.1){14}{\line(1,0){0.1}}
\multiput(3.9,-8.5)(0.15,-0.1){6}{\line(1,0){0.1}}
\multiput(10,-11)(-0.2,0.04){5}{\line(1,0){0.1}}
\put(10,-11){\line(1,0){2}}
\multiput(12,-11)(0.2,0.04){5}{\line(1,0){0.1}}
\multiput(19.5,-7.1)(0.1,0.15){6}{\line(1,0){0.1}}
\multiput(19.5,-7.1)(-0.1,-0.1){14}{\line(1,0){0.1}}
\multiput(18.1,-8.5)(-0.15,-0.1){6}{\line(1,0){0.1}}
\multiput(22,-1)(-0.04,-0.2){5}{\line(1,0){0.1}}
\put(22,0){\line(0,-1){1}}
\put(0,0){\circle*{3}}
\put(22,0){\circle*{3}}
\put(0,0){\line(1,0){22}}
\epi
+
\bpi(33,20)(-6,-3)
\put(2,0){\oval(4,4)[tl]}
\put(2,4){\oval(4,4)[br]}
\multiput(4,4)(-0.1,0.2){5}{\line(1,0){0.1}}
\multiput(3.5,5)(-0.1,0.1){5}{\line(1,0){0.1}}
\multiput(3,5.5)(-0.1,0.2){5}{\line(1,0){0.1}}
\put(4.5,6.5){\oval(4,4)[tl]}
\multiput(4.5,8.5)(0.2,-0.1){5}{\line(1,0){0.1}}
\multiput(5.5,8)(0.1,-0.1){5}{\line(1,0){0.1}}
\multiput(6,7.5)(0.2,-0.1){5}{\line(1,0){0.1}}
\put(7,9){\oval(4,4)[br]}
\put(11,9){\oval(4,4)[t]}
\put(15,9){\oval(4,4)[bl]}
\multiput(15,7)(0.2,0.1){5}{\line(1,0){0.1}}
\multiput(16,7.5)(0.1,0.1){5}{\line(1,0){0.1}}
\multiput(16.5,8)(0.2,0.1){5}{\line(1,0){0.1}}
\put(17.5,6.5){\oval(4,4)[tr]}
\multiput(19.5,6.5)(-0.1,-0.2){5}{\line(1,0){0.1}}
\multiput(19,5.5)(-0.1,-0.1){5}{\line(1,0){0.1}}
\multiput(18.5,5)(-0.1,-0.2){5}{\line(1,0){0.1}}
\put(20,4){\oval(4,4)[bl]}
\put(20,0){\oval(4,4)[tr]}
\put(2,0){\oval(4,4)[bl]}
\put(2,-4){\oval(4,4)[tr]}
\multiput(4,-4)(-0.1,-0.2){5}{\line(1,0){0.1}}
\multiput(3.5,-5)(-0.1,-0.1){5}{\line(1,0){0.1}}
\multiput(3,-5.5)(-0.1,-0.2){5}{\line(1,0){0.1}}
\put(4.5,-6.5){\oval(4,4)[bl]}
\multiput(4.5,-8.5)(0.2,0.1){5}{\line(1,0){0.1}}
\multiput(5.5,-8)(0.1,0.1){5}{\line(1,0){0.1}}
\multiput(6,-7.5)(0.2,0.1){5}{\line(1,0){0.1}}
\put(7,-9){\oval(4,4)[tr]}
\put(11,-9){\oval(4,4)[b]}
\put(15,-9){\oval(4,4)[tl]}
\multiput(15,-7)(0.2,-0.1){5}{\line(1,0){0.1}}
\multiput(16,-7.5)(0.1,-0.1){5}{\line(1,0){0.1}}
\multiput(16.5,-8)(0.2,-0.1){5}{\line(1,0){0.1}}
\put(17.5,-6.5){\oval(4,4)[br]}
\multiput(19.5,-6.5)(-0.1,0.2){5}{\line(1,0){0.1}}
\multiput(19,-5.5)(-0.1,0.1){5}{\line(1,0){0.1}}
\multiput(18.5,-5)(-0.1,0.2){5}{\line(1,0){0.1}}
\put(20,-4){\oval(4,4)[tl]}
\put(20,0){\oval(4,4)[br]}
\put(0,0){\circle*{3}}
\put(22,0){\circle*{3}}
\put(0,0){\line(1,0){22}}
\epi
+
\bpi(33,20)(-6,-3)
\put(0,0){\circle*{1}}
\put(0.8,4.2){\circle*{1}}
\put(3.2,7.8){\circle*{1}}
\put(6.8,10.2){\circle*{1}}
\put(11,11){\circle*{1}}
\put(15.2,10.2){\circle*{1}}
\put(18.8,7.8){\circle*{1}}
\put(21.2,4.2){\circle*{1}}
\put(22,0){\circle*{1}}
\multiput(10,11)(0.1,0.1){28}{\line(1,0){0.1}}
\multiput(10,11)(0.1,-0.1){28}{\line(1,0){0.1}}
\multiput(12,-11)(-0.1,0.1){28}{\line(1,0){0.1}}
\multiput(12,-11)(-0.1,-0.1){28}{\line(1,0){0.1}}
\put(0,0){\circle*{1}}
\put(0.8,-4.2){\circle*{1}}
\put(3.2,-7.8){\circle*{1}}
\put(6.8,-10.2){\circle*{1}}
\put(11,-11){\circle*{1}}
\put(15.2,-10.2){\circle*{1}}
\put(18.8,-7.8){\circle*{1}}
\put(21.2,-4.2){\circle*{1}}
\put(22,0){\circle*{1}}
\put(0,0){\circle*{3}}
\put(22,0){\circle*{3}}
\put(0,0){\line(1,0){22}}
\epi
+
\bpi(33,20)(-6,-3)
\put(11,1){\oval(22,22)[b]}
\put(0,0){\circle*{3}}
\put(22,0){\circle*{3}}
\multiput(3,0)(8,0){3}{\oval(4,4)[t]}
\multiput(7,0)(8,0){2}{\oval(4,4)[b]}
\put(0,0){\line(0,1){1}}
\multiput(0,1)(0.04,0.2){5}{\line(1,0){0.1}}
\multiput(2.5,7.1)(-0.1,-0.15){6}{\line(1,0){0.1}}
\multiput(2.5,7.1)(0.1,0.1){14}{\line(1,0){0.1}}
\multiput(3.9,8.5)(0.15,0.1){6}{\line(1,0){0.1}}
\multiput(10,11)(-0.2,-0.04){5}{\line(1,0){0.1}}
\put(10,11){\line(1,0){2}}
\multiput(12,11)(0.2,-0.04){5}{\line(1,0){0.1}}
\multiput(19.5,7.1)(0.1,-0.15){6}{\line(1,0){0.1}}
\multiput(19.5,7.1)(-0.1,0.1){14}{\line(1,0){0.1}}
\multiput(18.1,8.5)(-0.15,0.1){6}{\line(1,0){0.1}}
\multiput(22,1)(-0.04,0.2){5}{\line(1,0){0.1}}
\put(22,0){\line(0,1){1}}
\epi
\right.
\nn
\bpi(0,40)(0,0)
\epi
&&\left.\;\;\;\;
+
\bpi(33,30)(-6,-3)
\put(11,1){\circle{22}}
\put(11,13){\circle*{3}}
\put(11,13){\circle{5}}
\epi
+
\bpi(33,20)(-6,-3)
\put(0,0){\line(0,1){1}}
\multiput(0,1)(0.04,0.2){5}{\line(1,0){0.1}}
\multiput(2.5,7.1)(-0.1,-0.15){6}{\line(1,0){0.1}}
\multiput(2.5,7.1)(0.1,0.1){14}{\line(1,0){0.1}}
\multiput(3.9,8.5)(0.15,0.1){6}{\line(1,0){0.1}}
\multiput(10,11)(-0.2,-0.04){5}{\line(1,0){0.1}}
\put(10,11){\line(1,0){2}}
\multiput(12,11)(0.2,-0.04){5}{\line(1,0){0.1}}
\multiput(19.5,7.1)(0.1,-0.15){6}{\line(1,0){0.1}}
\multiput(19.5,7.1)(-0.1,0.1){14}{\line(1,0){0.1}}
\multiput(18.1,8.5)(-0.15,0.1){6}{\line(1,0){0.1}}
\multiput(22,1)(-0.04,0.2){5}{\line(1,0){0.1}}
\put(22,0){\line(0,1){1}}
\put(0,0){\line(0,-1){1}}
\multiput(0,-1)(0.04,-0.2){5}{\line(1,0){0.1}}
\multiput(2.5,-7.1)(-0.1,0.15){6}{\line(1,0){0.1}}
\multiput(2.5,-7.1)(0.1,-0.1){14}{\line(1,0){0.1}}
\multiput(3.9,-8.5)(0.15,-0.1){6}{\line(1,0){0.1}}
\multiput(10,-11)(-0.2,0.04){5}{\line(1,0){0.1}}
\put(10,-11){\line(1,0){2}}
\multiput(12,-11)(0.2,0.04){5}{\line(1,0){0.1}}
\multiput(19.5,-7.1)(0.1,0.15){6}{\line(1,0){0.1}}
\multiput(19.5,-7.1)(-0.1,-0.1){14}{\line(1,0){0.1}}
\multiput(18.1,-8.5)(-0.15,-0.1){6}{\line(1,0){0.1}}
\multiput(22,-1)(-0.04,-0.2){5}{\line(1,0){0.1}}
\put(22,0){\line(0,-1){1}}
\put(11,11){\circle*{3}}
\put(11,11){\circle{5}}
\epi
+
\bpi(33,20)(-6,-3)
\put(2,0){\oval(4,4)[tl]}
\put(2,4){\oval(4,4)[br]}
\multiput(4,4)(-0.1,0.2){5}{\line(1,0){0.1}}
\multiput(3.5,5)(-0.1,0.1){5}{\line(1,0){0.1}}
\multiput(3,5.5)(-0.1,0.2){5}{\line(1,0){0.1}}
\put(4.5,6.5){\oval(4,4)[tl]}
\multiput(4.5,8.5)(0.2,-0.1){5}{\line(1,0){0.1}}
\multiput(5.5,8)(0.1,-0.1){5}{\line(1,0){0.1}}
\multiput(6,7.5)(0.2,-0.1){5}{\line(1,0){0.1}}
\put(7,9){\oval(4,4)[br]}
\put(11,9){\oval(4,4)[t]}
\put(15,9){\oval(4,4)[bl]}
\multiput(15,7)(0.2,0.1){5}{\line(1,0){0.1}}
\multiput(16,7.5)(0.1,0.1){5}{\line(1,0){0.1}}
\multiput(16.5,8)(0.2,0.1){5}{\line(1,0){0.1}}
\put(17.5,6.5){\oval(4,4)[tr]}
\multiput(19.5,6.5)(-0.1,-0.2){5}{\line(1,0){0.1}}
\multiput(19,5.5)(-0.1,-0.1){5}{\line(1,0){0.1}}
\multiput(18.5,5)(-0.1,-0.2){5}{\line(1,0){0.1}}
\put(20,4){\oval(4,4)[bl]}
\put(20,0){\oval(4,4)[tr]}
\put(2,0){\oval(4,4)[bl]}
\put(2,-4){\oval(4,4)[tr]}
\multiput(4,-4)(-0.1,-0.2){5}{\line(1,0){0.1}}
\multiput(3.5,-5)(-0.1,-0.1){5}{\line(1,0){0.1}}
\multiput(3,-5.5)(-0.1,-0.2){5}{\line(1,0){0.1}}
\put(4.5,-6.5){\oval(4,4)[bl]}
\multiput(4.5,-8.5)(0.2,0.1){5}{\line(1,0){0.1}}
\multiput(5.5,-8)(0.1,0.1){5}{\line(1,0){0.1}}
\multiput(6,-7.5)(0.2,0.1){5}{\line(1,0){0.1}}
\put(7,-9){\oval(4,4)[tr]}
\put(11,-9){\oval(4,4)[b]}
\put(15,-9){\oval(4,4)[tl]}
\multiput(15,-7)(0.2,-0.1){5}{\line(1,0){0.1}}
\multiput(16,-7.5)(0.1,-0.1){5}{\line(1,0){0.1}}
\multiput(16.5,-8)(0.2,-0.1){5}{\line(1,0){0.1}}
\put(17.5,-6.5){\oval(4,4)[br]}
\multiput(19.5,-6.5)(-0.1,0.2){5}{\line(1,0){0.1}}
\multiput(19,-5.5)(-0.1,0.1){5}{\line(1,0){0.1}}
\multiput(18.5,-5)(-0.1,0.2){5}{\line(1,0){0.1}}
\put(20,-4){\oval(4,4)[tl]}
\put(20,0){\oval(4,4)[br]}
\put(11,11){\circle*{3}}
\put(11,11){\circle{5}}
\epi
+
\bpi(33,20)(-6,-3)
\put(0,0){\circle*{1}}
\put(0.8,4.2){\circle*{1}}
\put(3.2,7.8){\circle*{1}}
\put(6.8,10.2){\circle*{1}}
\put(11,11){\circle*{1}}
\put(15.2,10.2){\circle*{1}}
\put(18.8,7.8){\circle*{1}}
\put(21.2,4.2){\circle*{1}}
\put(22,0){\circle*{1}}
\multiput(12,-11)(-0.1,0.1){28}{\line(1,0){0.1}}
\multiput(12,-11)(-0.1,-0.1){28}{\line(1,0){0.1}}
\put(0,0){\circle*{1}}
\put(0.8,-4.2){\circle*{1}}
\put(3.2,-7.8){\circle*{1}}
\put(6.8,-10.2){\circle*{1}}
\put(11,-11){\circle*{1}}
\put(15.2,-10.2){\circle*{1}}
\put(18.8,-7.8){\circle*{1}}
\put(21.2,-4.2){\circle*{1}}
\put(22,0){\circle*{1}}
\put(11,11){\circle*{3}}
\put(11,11){\circle{5}}
\epi
+
\bpi(33,20)(-6,-3)
\put(11,0){\circle*{3}}
\put(11,0){\circle{5}}
\put(11,0){\circle{8}}
\epi
\right]
\nn
&=&
\frac{g^2}{(4\pi)^4m_A^2}
\{-\threequarter m_{HG\gh}^4K_{HHH}-\quarter(m_{HG\gh}^2
+2m_{\gh}^2)^2K_{HGG}
\nn
&&-\quarter[(m_H^2-2m_A^2)^2+8m_A^2]K_{H\!AA}
-\quarter[(m_H^2-2m_{\gh}^2)^2-8m_{\gh}^2]K_{H\gh\gh}
\nn
&&+\half T_{H\!A\gh}K_{H\!A\gh}-\half T_{HGA}K_{HGA}
+\half m_{HG\gh}^4K_{HG\gh}
\nn
&&+\eighth m_{HG\gh}^2[3(L_H^2+L_G^2)+2L_HL_G]
+\half m_A^2[L_HL_G+2(L_H+L_G)L_A]
\nn
&&+\quarter m_H^2(L_A-L_{\gh})^2
\nn
&&+\half[(m_A^2-m_{HG\gh}^2)L_H+(m_H^2-m_G^2)L_G
-(m_A^2L_A-m_{\gh}^2L_{\gh})](L_A-L_{\gh})
\nn
&&+m_A^4(3L_H+L_G)+m_A^2(m_H^2+m_G^2+\twothird m_A^2-m_4^2)L_A
\nn
&&-m_A^4(m_H^2+2m_A^2)\}\,,
\eea
where
\bea
T_{xyz}&\equiv&m_x^4+m_y^4+m_z^4-2m_y^2m_z^2-2m_z^2m_x^2-2m_x^2m_y^2\,,
\\
m_{HG\gh}^2&\equiv&m_H^2-m_G^2+m_{\gh}^2=2\la\vp^2,
\\
L_x&\equiv&m_x^2[\ln(m_x^2/\bar{\mu}^2)-1],
\\
K_{xyz}
&\equiv&K(m_x^2,m_y^2,m_z^2)
\nn
&\equiv&\lim_{\ep\rightarrow 0}\left\{(4\pi)^4
\int\frac{d^dp}{(2\pi)^d}\frac{d^dq}{(2\pi)^d}
\frac{(\mu^2)^\ep}{(p^2-m_x^2)(q^2-m_y^2)[(p+q)^2-m_z^2]}\right.
\nn
&&\left.+\sum_{n=x}^zm_n^2\left[\frac{2}{\ep^2}
-\frac{2}{\ep}\left(\ln\frac{m_n^2}{\bar{\mu}^2}-\frac{3}{2}\right)
+\frac{1}{2}\left(\ln\frac{m_n^2}{\bar{\mu}^2}-1\right)^2
+\frac{\pi^2+6}{12}\right]\right\}.
\nn
\eea
(For an evaluation of $K_{xyz}$ see e.g.\ \cite{FoJa..}.)
$V_{1,\overline{R}_\xi}$ and $V_{2,\overline{R}_\xi}$ are only slightly more
complicated than the corresponding expressions in Landau gauge which one gets
back by letting $\xi\rightarrow 0$ in our results.

Since the EP is not itself a physical quantity it can be and indeed is
gauge dependent. However, its value at points where $V'=0$ is a physical
energy density and should therefore be gauge independent \cite{Ni,AiFr,FuKu}.
The energy density at the symmetry breaking solution of $V'=0$ is given by
\beq
\label{V(v)}
V(v) = i\left.\left\{
\bpi(15,20)(0,16.5)
\put(7,20){\circle*{3}}
\epi
+
\bpi(44,20)(0,16.5)
\put(22,20){\circle{32}}
\put(22,21){\makebox(0,0)[b]{\footnotesize 1-loop}}
\put(22,17){\makebox(0,0)[t]{\mbox{\footnotesize 1PI}}}
\epi
+\left[
\bpi(44,20)(0,16.5)
\put(22,20){\circle{32}}
\put(22,21){\makebox(0,0)[b]{\footnotesize 2-loop}}
\put(22,17){\makebox(0,0)[t]{\mbox{\footnotesize 1PI}}}
\epi
+
\bpi(92,20)(0,16.5)
\put(22,20){\circle{32}}
\put(22,21){\makebox(0,0)[b]{\footnotesize 1-loop}}
\put(22,17){\makebox(0,0)[t]{\mbox{\footnotesize 1PI}}}
\put(38,20){\line(1,0){16}}
\put(70,20){\circle{32}}
\put(70,21){\makebox(0,0)[b]{\footnotesize 1-loop}}
\put(70,17){\makebox(0,0)[t]{\mbox{\footnotesize 1PI}}}
\epi
\right]
+
\begin{array}{cc}{\rm higher}\\{\rm loops}\end{array}
\right\}\right|_{\vp=v_0},
\eeq
where
\beq
\bpi(59,36)(0,0)
\put(22,20){\circle{32}}
\put(22,21){\makebox(0,0)[b]{\footnotesize 1-loop}}
\put(22,17){\makebox(0,0)[t]{\mbox{\footnotesize 1PI}}}
\put(22,-12){\line(0,1){16}}
\epi
=
\bpi(44,11)(-11,-15)
\put(11,1){\circle{22}}
\put(11,-27){\line(0,1){16}}
\put(11,-11){\circle*{3}}
\epi
+
\bpi(44,11)(-11,-15)
\put(0,0){\line(0,1){1}}
\multiput(0,1)(0.04,0.2){5}{\line(1,0){0.1}}
\multiput(2.5,7.1)(-0.1,-0.15){6}{\line(1,0){0.1}}
\multiput(2.5,7.1)(0.1,0.1){14}{\line(1,0){0.1}}
\multiput(3.9,8.5)(0.15,0.1){6}{\line(1,0){0.1}}
\multiput(10,11)(-0.2,-0.04){5}{\line(1,0){0.1}}
\put(10,11){\line(1,0){2}}
\multiput(12,11)(0.2,-0.04){5}{\line(1,0){0.1}}
\multiput(19.5,7.1)(0.1,-0.15){6}{\line(1,0){0.1}}
\multiput(19.5,7.1)(-0.1,0.1){14}{\line(1,0){0.1}}
\multiput(18.1,8.5)(-0.15,0.1){6}{\line(1,0){0.1}}
\multiput(22,1)(-0.04,0.2){5}{\line(1,0){0.1}}
\put(22,0){\line(0,1){1}}
\put(0,0){\line(0,-1){1}}
\multiput(0,-1)(0.04,-0.2){5}{\line(1,0){0.1}}
\multiput(2.5,-7.1)(-0.1,0.15){6}{\line(1,0){0.1}}
\multiput(2.5,-7.1)(0.1,-0.1){14}{\line(1,0){0.1}}
\multiput(3.9,-8.5)(0.15,-0.1){6}{\line(1,0){0.1}}
\multiput(10,-11)(-0.2,0.04){5}{\line(1,0){0.1}}
\put(10,-11){\line(1,0){2}}
\multiput(12,-11)(0.2,0.04){5}{\line(1,0){0.1}}
\multiput(19.5,-7.1)(0.1,0.15){6}{\line(1,0){0.1}}
\multiput(19.5,-7.1)(-0.1,-0.1){14}{\line(1,0){0.1}}
\multiput(18.1,-8.5)(-0.15,-0.1){6}{\line(1,0){0.1}}
\multiput(22,-1)(-0.04,-0.2){5}{\line(1,0){0.1}}
\put(22,0){\line(0,-1){1}}
\put(11,-27){\line(0,1){16}}
\put(11,-11){\circle*{3}}
\epi
+
\bpi(44,11)(-11,-15)
\put(2,0){\oval(4,4)[tl]}
\put(2,4){\oval(4,4)[br]}
\multiput(4,4)(-0.1,0.2){5}{\line(1,0){0.1}}
\multiput(3.5,5)(-0.1,0.1){5}{\line(1,0){0.1}}
\multiput(3,5.5)(-0.1,0.2){5}{\line(1,0){0.1}}
\put(4.5,6.5){\oval(4,4)[tl]}
\multiput(4.5,8.5)(0.2,-0.1){5}{\line(1,0){0.1}}
\multiput(5.5,8)(0.1,-0.1){5}{\line(1,0){0.1}}
\multiput(6,7.5)(0.2,-0.1){5}{\line(1,0){0.1}}
\put(7,9){\oval(4,4)[br]}
\put(11,9){\oval(4,4)[t]}
\put(15,9){\oval(4,4)[bl]}
\multiput(15,7)(0.2,0.1){5}{\line(1,0){0.1}}
\multiput(16,7.5)(0.1,0.1){5}{\line(1,0){0.1}}
\multiput(16.5,8)(0.2,0.1){5}{\line(1,0){0.1}}
\put(17.5,6.5){\oval(4,4)[tr]}
\multiput(19.5,6.5)(-0.1,-0.2){5}{\line(1,0){0.1}}
\multiput(19,5.5)(-0.1,-0.1){5}{\line(1,0){0.1}}
\multiput(18.5,5)(-0.1,-0.2){5}{\line(1,0){0.1}}
\put(20,4){\oval(4,4)[bl]}
\put(20,0){\oval(4,4)[tr]}
\put(2,0){\oval(4,4)[bl]}
\put(2,-4){\oval(4,4)[tr]}
\multiput(4,-4)(-0.1,-0.2){5}{\line(1,0){0.1}}
\multiput(3.5,-5)(-0.1,-0.1){5}{\line(1,0){0.1}}
\multiput(3,-5.5)(-0.1,-0.2){5}{\line(1,0){0.1}}
\put(4.5,-6.5){\oval(4,4)[bl]}
\multiput(4.5,-8.5)(0.2,0.1){5}{\line(1,0){0.1}}
\multiput(5.5,-8)(0.1,0.1){5}{\line(1,0){0.1}}
\multiput(6,-7.5)(0.2,0.1){5}{\line(1,0){0.1}}
\put(7,-9){\oval(4,4)[tr]}
\put(11,-9){\oval(4,4)[b]}
\put(15,-9){\oval(4,4)[tl]}
\multiput(15,-7)(0.2,-0.1){5}{\line(1,0){0.1}}
\multiput(16,-7.5)(0.1,-0.1){5}{\line(1,0){0.1}}
\multiput(16.5,-8)(0.2,-0.1){5}{\line(1,0){0.1}}
\put(17.5,-6.5){\oval(4,4)[br]}
\multiput(19.5,-6.5)(-0.1,0.2){5}{\line(1,0){0.1}}
\multiput(19,-5.5)(-0.1,0.1){5}{\line(1,0){0.1}}
\multiput(18.5,-5)(-0.1,0.2){5}{\line(1,0){0.1}}
\put(20,-4){\oval(4,4)[tl]}
\put(20,0){\oval(4,4)[br]}
\put(11,-27){\line(0,1){16}}
\put(11,-11){\circle*{3}}
\epi
+
\bpi(44,11)(-11,-15)
\put(0,0){\circle*{1}}
\put(0.8,4.2){\circle*{1}}
\put(3.2,7.8){\circle*{1}}
\put(6.8,10.2){\circle*{1}}
\put(11,11){\circle*{1}}
\put(15.2,10.2){\circle*{1}}
\put(18.8,7.8){\circle*{1}}
\put(21.2,4.2){\circle*{1}}
\put(22,0){\circle*{1}}
\multiput(10,11)(0.1,0.1){28}{\line(1,0){0.1}}
\multiput(10,11)(0.1,-0.1){28}{\line(1,0){0.1}}
\put(0,0){\circle*{1}}
\put(0.8,-4.2){\circle*{1}}
\put(3.2,-7.8){\circle*{1}}
\put(6.8,-10.2){\circle*{1}}
\put(11,-11){\circle*{1}}
\put(15.2,-10.2){\circle*{1}}
\put(18.8,-7.8){\circle*{1}}
\put(21.2,-4.2){\circle*{1}}
\put(22,0){\circle*{1}}
\put(11,-27){\line(0,1){16}}
\put(11,-11){\circle*{3}}
\epi
+
\bpi(50,11)(-11,-15)
\put(11,0){\circle*{3}}
\put(11,0){\circle{5}}
\put(11,-27){\line(0,1){27}}
\epi
{}.
\eeq
The one-loop contribution is gotten by substituting $v_0$ into
(\ref{v1rxi}) and (\ref{v1rxibar}) and one finds
$V_{1,\overline{R}_\xi}(v_0)=V_{1,R_\xi}(v_0)=$
gauge independent as expected although the one-loop correction to the Higgs
field vev, given by $v_1=V_1'(v_0)/(2m^2)$, turns out to be gauge dependent
as is expected for the location of the vev \cite{Ni}. In $\overline{R}_\xi$
gauge it is easy to determine the two-loop contribution to (\ref{V(v)}) and
it also turns out to be gauge independent. The same is true for the
other stationary point, i.e.\ $\vp=0$ (however, the loop expansion
is a bad approximation scheme here due to infrared divergences).

\section{Application II: An SU(2)-Higgs Model}
\label{su2higgs}
For the Abelian Higgs model $R_{abc}$ and $S_{abc}$ vanish. To demonstrate
some non-trivial appearance of $R_{abc}$ the renormalization of an
SU(2)-Higgs model will be sketched in this section. This model can also
be regarded as a truncated version of the standard electroweak model
with vanishing weak mixing angle and no fermions.

We start from the Lagrangian
\beq
\label{nahi}
\cL = \frac{1}{2}\left(D_\mu\Phi\right)^T\left(D^\mu\Phi\right)
-\frac{1}{4}F_{a\mu\nu}F_a^{\mu\nu}-\frac{1}{2}m^2\Phi^T\Phi
-\frac{\la}{4}(\Phi^T\Phi)^2
\eeq
with
\bea
\label{fmunu}
F_a^{\mu\nu} &=& \partial^\mu A_a^\nu-\partial^\nu A_a^\mu
-g\ep_{abc}A_b^\mu A_c^\nu\,,
\\
\label{dmuphi}
D^\mu\Phi &=& \left(\partial^\mu+igT_aA_a^\mu\right)\Phi\,.
\eea
Here
\beq
\label{phit4}
\Phi^T=(\phi_0,\phi_1,\phi_2,\phi_3)
\eeq
is a collection of real scalar fields and with the scalar
self-coupling $\la>0$ and $m^2<0$ the SU(2) gauge symmetry
is spontaneously broken down completely.

It turns out to be clever to choose the generators as
\beq
T_{cAB}=-\frac{i}{2}\et_{ABc}\,,
\eeq
where the $\et_{ABc}$ are 't Hooft symbols \cite{tHeta}
\beq
\begin{array}{lcrcrcl}
\et_{ABc}&=&\ep_{ABc}&{\rm for}&A,B,c&=&1,2,3\\
\et_{A0c}&=&\delta_{Ac}&{\rm for}&A,c&=&1,2,3\\
\et_{0Bc}&=&-\delta_{Bc}&{\rm for}&B,c&=&1,2,3\\
\et_{00c}&=&0&{\rm for}&c&=&1,2,3\,,\\
\end{array}
\eeq
and the direction of symmetry breaking as
\beq
\vh=(1,0,0,0).
\eeq
Then $\Phi$ is nothing but a complex doublet transforming under the
fundamental representation of SU(2), written in terms of its real
components. The normalization is such that $f_{abc}=\ep_{abc}$. Since
there is no unbroken subgroup, we try
\beq
\sg_{ab}=\sg\de_{ab}/\xi,\;\;\;\;\;\;\;\;\;\;\;\;\Th_{ab}=\xi\de_{ab}\,.
\eeq
{}From the structure of the one-loop divergences for
the quartic ghost coupling for $R_{abc}=S_{abc}=0$ it is then easy to guess
as simplest admissible form
\beq
R_{abc}=\al\xi\ep_{abc}\,,\;\;\;\;\;\;\;\;\;\;\;\;S_{abc}=0\,.
\eeq
By considering all divergent one-loop diagrams we can now check
renormalizability to this order. Indeed everything works out and with
\beq
\phi_{0B}=Z_H^{\frac{1}{2}}\phi_{0R}\,,\;\;\;
\phi_{aB}=Z_G^{\frac{1}{2}}\phi_{aR}\,,\;\;\;
\al_B=Z_\al\al_R\,,
\eeq
$a=1,2,3$, and otherwise the same definitions as in (\ref{zx}) the resulting
effective Lagrangian $\cL_\rmeff=\cL+\cL_\rmgf+\cL_\rmgh$ keeps
its form under renormalization. The one-loop results for the $Z_x$ can
be found in appendix B. To enjoy the absence of mixed
would-be-Goldstone-gauge propagators we would set $\sg_R=1$.

$\cL_\rmeff$ can also be obtained by the following procedure: Take the
most general ${\rm dim}\leq 4$ Lagrangian with given field content
$(\phi_0,\phi_a,A_a^\mu,\eb_a,\et_a,B_a)$, $a=1,2,3$. Impose the
nil-potent $s$-symmetry of (\ref{brst}) with $T_a$ and $f_{abc}$ given
above, the discrete symmetry $(\phi_0,\phi_a)\rightarrow(-\phi_0,-\phi_a)$
and a global SO(3) symmetry, under which $\phi_0$ is a scalar and $\phi_a$,
$A_a^\mu$, $\eb_a$, $\et_a$, $B_a$ are vectors. Integrate out the $B_a$. Up
to total divergencies and trivial changes of variables, $\cL_\rmeff$ with
the parameters given above is the result. Therefore $\cL_\rmeff$ is
renormalizable in any regularization scheme that observes the symmetries,
e.g.\ dimensional regularization.

To impose $\sbar$-invariance for this case we define
\beq
\De_{ab}=\xi\de_{ab}
\eeq
so that (\ref{theta}) is fulfilled. Using (\ref{rabc}) and (\ref{sabc})
it follows
\beq
R_{abc}=-\frac{\xi}{2\sg}\ep_{abc}\,,\;\;\;\;\;\;\;\;\;\;\;\;S_{abc}=0
\eeq
and therefore $\sbar$-invariance is equivalent to
\beq
\al=-\frac{1}{2\sg}\,.
\eeq
Stability of this condition under renormalization can easily be checked
at the one-loop level using the results for $Z_\sg$ and $Z_\al$ given in
appendix B.

\section{Application III: An SO(3)-Higgs Model}
\label{so3higgs}
For both examples considered so far, $\sg_{ab}$ is proportional to
$\de_{ab}$ and $S_{abc}=0$. To present a non-trivial appearance of
both $\sg_{ab}$ and $S_{abc}$, now the renormalization of an SO(3)-Higgs
model with an unbroken SO(2) subgroup will be sketched.

Consider again the Lagrangian (\ref{nahi}), with $F_a^{\mu\nu}$ and
$D^\mu\Phi$ as in (\ref{fmunu}) and (\ref{dmuphi}), but now with a
triplet of real scalar fields
\beq
\label{phit3}
\Phi^T=(\phi_1,\phi_2,\phi_3)
\eeq
and the generators given by
\beq
T_{ijk}=-i\ep_{ijk}\,.
\eeq
Again the normalization is such that $f_{abc}=\ep_{abc}$.

With the direction of symmetry breaking chosen as
\beq
\vh=(0,0,1)
\eeq
there is an unbroken SO(2) subgroup of rotations in field space around the
3-axis. While $\Th_{ab}$ contributes only for broken generators and we can
set
\beq
\Th_{ab}=\xi\de_{ab},
\eeq
we have to treat broken and unbroken generators differently in $\sg_{ab}$,
$R_{abc}$ and $S_{abc}$. For $\sg_{ab}$ it is natural to try
\beq
\sg_{ab}=\frac{\sw}{\xi}(\de_{ab}-\de_{a3}\de_{b3})
+\frac{\sa}{\xi}\de_{a3}\de_{b3}.
\eeq
In order to minimize the amount of complication involved we choose
$R_{abc}$ and $S_{abc}$ to be non-zero only if $\{a,b,c\}$ is a
permutation of $\{1,2,3\}$. Together with the symmetry properties
(\ref{rssym}) we get therefore
\bea
R_{abc}&=&\xi\al\ep_{ab3}\de_{3c}+\xi\be(\de_{a3}\ep_{3bc}-\de_{b3}\ep_{3ac}),
\\
S_{abc}&=&\half\xi\ga(\ep_{ab3}\de_{3c}+\ep_{ac3}\de_{3b}).
\eea
Considering again all divergent one-loop diagrams and with
\beq
\!\!
\begin{array}{c}
\phi_{1,2B}=Z_G^{\frac{1}{2}}\phi_{1,2R}\,,\;\;\;
\phi_{3B}=Z_H^{\frac{1}{2}}\phi_{3R}\,,\;\;\;
A_{1,2B}^\mu=Z_W^\frac{1}{2}A_{1,2R}^\mu\,,\;\;\;
A_{3B}^\mu=Z_A^\frac{1}{2}A_{3R}^\mu\,,
\vsd\\
\eb_{1,2B}=Z_{\eb_W}\eb_{1,2R}\,,\;\;\;
\eb_{3B}=Z_{\eb_A}\eb_{3R}\,,\;\;\;
\et_{1,2B}=Z_{\et_W}\et_{1,2R}\,,\;\;\;
\et_{3B}=Z_{\et_A}\et_{3R}\,,
\vsd\\
\al_B=Z_\al\al_R\,,\;
\be_B=Z_\be\be_R\,,\;
\ga_B=Z_\ga\ga_R\,,\;
\sg_{W\!B}=Z_{\sw}\sg_{W\!R}\,,\;
\sg_{AB}=Z_{\sa}\sg_{AR}\,,
\end{array}
\eeq
and otherwise the same definitions as in (\ref{zx}) the resulting
effective Lagrangian keeps its form under renormalization. The
one-loop results for the $Z_x$ can be found in appendix C. To enjoy
the absence of mixed would-be-Goldstone-gauge propagators we would
set $\sg_{W\!R}=1$.

$\cL_\rmeff$ can also be obtained by the following procedure: Take the
most general ${\rm dim}\leq 4$ Lagrangian with given field content
$\psi_a\equiv(\phi_a,A_a^\mu,\eb_a,\et_a,B_a)$, $a=1,2,3$. Impose the
nil-potent $s$-symmetry of (\ref{brst}) with $T_a$ and $f_{abc}$ given
above, the discrete symmetries
$(\phi_1,\phi_2,\phi_3)\rightarrow(-\phi_1,-\phi_2,-\phi_3)$ and
$(\psi_1,\psi_3)\rightarrow(-\psi_1,-\psi_3)$ and a global SO(2) symmetry,
under which $\psi_3$ are scalars and $\psi_a$, $a=1,2$ are vectors.
Integrate out the $B_a$, $a=1,2,3$. Up to total divergencies and trivial
changes of variables, $\cL_\rmeff$ with the parameters given above is the
result. Therefore $\cL_\rmeff$ is renormalizable in any regularization
scheme that observes the symmetries, e.g.\ dimensional regularization.

To impose $\sbar$-invariance for this case we define
\beq
\De_{ab}=\xi(\de_{ab}-\de_{a3}\de_{b3})+\xi\Da^{-1}\de_{a3}\de_{b3}
\eeq
so that (\ref{theta}) is fulfilled. Using (\ref{rabc}) and (\ref{sabc})
it follows that
\bea
R_{abc}&=&\left(\frac{\Da^2}{2\sa}-\frac{1}{\sw}\right)\xi\ep_{ab3}\de_{c3}
-\frac{\Da}{2\sa}\xi(\de_{a3}\ep_{bc3}-\de_{b3}\ep_{ac3}),\\
S_{abc}&=&\half(1-\Da)(\ep_{ab3}\de_{c3}+\ep_{ac3}\de_{b3})
\eea
and therefore
\beq
\al=\frac{\Da^2}{2\sa}-\frac{1}{\sw}\,,\;\;\;\;\;\;\;\;\;\;\;
\be=-\frac{\Da}{2\sa}\,,\;\;\;\;\;\;\;\;\;\;\;
\ga=\frac{1-\Da}{\xi}\,.
\eeq
Eliminating the newly introduced parameter $\Da$ from these equations we get
\beq
\label{albe}
\al=\frac{(\ga\xi-1)^2}{2\sa}-\frac{1}{\sw}\,,\;\;\;\;\;\;\;\;\;\;\;\;
\be=\frac{\ga\xi-1}{2\sa}\,.
\eeq

In the case at hand we can diminish the number of gauge parameters by
imposing
\beq
\label{beeq0}
\be=0
\eeq
alternatively or additionally to $\sbar$-invariance.
This condition can be shown to be stable under renormalization.
Note that if we impose (\ref{beeq0}) additionally to $s$- and
$\sbar$-symmetry, $\eb_3$ and $\et_3$ effectively drop from the
theory. As already noted in section \ref{ssbar} this is due to
the fact that an unbroken Abelian subgroup does not always
require a ghost field. Also, since now $\Da=0$, we loose invertibility
of $\De_{ab}^{-1}$; only the restriction of $\De_{ab}^{-1}$ to
$a,b=1,2$ can be inverted which turns out to be sufficient at this stage.
We have
\beq
\label{albeandbeeq0}
\al=-\frac{1}{\sw}\,,\;\;\;\;\;\;\;\;\;\;\;\;
\be=0,\;\;\;\;\;\;\;\;\;\;\;\;\ga=\frac{1}{\xi}\,,
\eeq
which also means that $S_{abc}$ is no longer renormalized.
Stability of (\ref{albe}), (\ref{beeq0}) and (\ref{albeandbeeq0})
under renormalization can easily be checked at the one-loop level using
the results given in appendix C.

\vsb

I am grateful to Roberto Peccei, Duncan Morris, Gerard \mbox{'t Hooft},
Konstadinos Sfetsos and especially Bernard de Wit for valuable discussions
and suggestions, to Jim Congleton for proofreading an early version
of the manuscript and to the referee for crucial suggestions. This work was
supported by Stichting FOM.

\appendix
\section*{Appendix A}
Using dimensional regularization and the MS scheme
the $Z_x$ up to two loops in the Abelian Higgs model are for $\sg=1$
(all quantities are the renormalized ones, $\ep=4-d$, $d=$ dimension of
space-time):
\begin{eqnarray}
\mu^\ep Z_H &=& 1+\frac{(6+2\xi)g^2}{(4\pi)^2\ep}
+\frac{-4\la^2-4\la\xi g^2+(-\frac{10}{3}+2\xi-3\xi^2)g^4}
{(4\pi)^4\ep}
\nn
&&+\frac{8\la\xi g^2+(20+12\xi^2)g^4}{(4\pi)^4\ep^2}
\\
\mu^\ep Z_G &=& 1+\frac{(6-6\xi)g^2}{(4\pi)^2\ep}
+\frac{-4\la^2+4\la\xi g^2+(-\frac{10}{3}-2\xi-11\xi^2)g^4}
{(4\pi)^4\ep}
\nn
&&+\frac{-8\la\xi g^2+(20-24\xi+12\xi^2)g^4}{(4\pi)^4\ep^2}
\\
\mu^\ep Z_A &=& 1-\frac{\frac{2}{3}g^2}{(4\pi)^2\ep}-\frac{4g^4}{(4\pi)^4\ep}
\\
\mu^{\ep}Z_\et &=& 1-\frac{2\xi^2 g^4}{(4\pi)^4\ep}
\\
Z_m &=& 1+\frac{8\la-6g^2}{(4\pi)^2\ep}
+\frac{-20\la^2+32\la g^2+\frac{43}{3}g^4}{(4\pi)^4\ep}
\nn
&&+\frac{112\la^2-96\la g^2+40g^4}{(4\pi)^4\ep^2}
\\
\mu^{-\ep}Z_\la &=& 1+\frac{20\la-12g^2+6g^4/\la}{(4\pi)^2\ep}
+\frac{-120\la^2+56\la g^2+\frac{158}{3}g^4
-\frac{104}{3}g^6/\la}{(4\pi)^4\ep}
\nn
&&+\frac{400\la^2-360\la g^2+188g^4-32g^6/\la}{(4\pi)^4\ep^2}
\\
\mu^{-\ep}Z_g &=& 1+\frac{\frac{2}{3}g^2}{(4\pi)^2\ep}
+\frac{4g^4}{(4\pi)^4\ep}+\frac{\frac{4}{9}g^4}{(4\pi)^4\ep^2}
\\
Z_\xi &=& 1+\frac{4\la+(-\frac{20}{3}+4\xi)g^2}{(4\pi)^2\ep}
+\frac{-12\la^2-8\la g^2+(\frac{1}{3}+12\xi^2)g^4}{(4\pi)^4\ep}
\nn
&&+\frac{48\la^2+(-\frac{152}{3}+16\xi)\la g^2
+(32-\frac{80}{3}\xi+4\xi^2)g^4}{(4\pi)^4\ep^2}
\\
Z_\sg &=& 1+\frac{4\la+(-6+6\xi)g^2}{(4\pi)^2\ep}
+\frac{-12\la^2-8\la g^2+(\frac{13}{3}-4\xi+12\xi^2)g^4}{(4\pi)^4\ep}
\nn
&&+\frac{48\la^2+(-48+32\xi)\la g^2+(28-48\xi+24\xi^2)g^4}{(4\pi)^4\ep^2}
\end{eqnarray}
Note that $Z_gZ_A=1$ up to terms of higher than two-loop order as
required by the Ward identity.

The one- and two-loop counterterms can be reconstructed from the $Z_x$ above.
For illustrative purposes and because some of them are used in the text,
the one-loop counterterms are given in table
\ref{counterterms1}.
\begin{table}[t]
\begin{tabular}{c|c}
\bpi(0,30)(0,-10)\put(0,3){\circle*{3}}\put(0,3){\circle{5}}\epi
&
\raisebox{10pt}{
$\displaystyle
\frac{-i}{2(4\pi)^2\ep}\left[
\left(10\la^2+2\la\xi g^2+3g^4\right)\vp^4
+\left(8\la+2\xi g^2\right)m^2\vp^2+2m^4
\right]$
}
\\
\hline
\bpi(0,30)(0,-10)
\put(0,10){\circle*{3}}
\put(0,10){\circle{5}}
\put(0,-5){\line(0,1){15}}
\epi
&
\raisebox{10pt}{
$\displaystyle \frac{-2i}{(4\pi)^2\ep}\left[
\left(10\la^2+2\la\xi g^2+3g^4\right)\vp^3
+\left(4\la+\xi g^2\right)m^2\vp
\right]$
}
\\
\hline
\bpi(36,48)(0,-20)
\put(0,3){\line(1,0){36}}
\put(18,3){\circle*{3}}
\put(18,3){\circle{5}}
\epi
&
\raisebox{20pt}{
$\begin{array}{rl}
\displaystyle\frac{i}{(4\pi)^2\ep}\left[\invexp\right.
\!\!\!\!\!&
(6+2\xi)g^2 k^2\\
&-(60\la^2+12\la\xi g^2+18g^4)\vp^2
-(8\la+2\xi g^2)m^2\left.\invexp\right]
\end{array}$
}
\\
\hline
\bpi(36,48)(0,-20)
\multiput(0,3)(8,0){5}{\line(1,0){4}}
\put(18,3){\circle*{3}}
\put(18,3){\circle{5}}
\epi
&
\raisebox{20pt}{
$\begin{array}{rl}
\displaystyle\frac{i}{(4\pi)^2\ep}\left[\invexp\right.
\!\!\!\!\!&
6(1-\xi)g^2 k^2\\
&-(20\la^2+4\la\xi g^2+6g^4+6\xi^2g^4)\vp^2
-(8\la-6\xi g^2)m^2\left.\invexp\right]
\end{array}$
}
\\
\hline
\bpi(48,30)(-8,-10)
\put(-9,0){$\mu$}
\multiput(2,3)(8,0){2}{\oval(4,4)[t]}
\multiput(6,3)(8,0){2}{\oval(4,4)[b]}
\multiput(16,3)(8,0){3}{\line(1,0){4}}
\put(16,3){\circle*{3}}
\put(16,3){\circle{5}}
\multiput(28,3)(-0.1,0.1){30}{\line(1,0){0.1}}
\multiput(28,3)(-0.1,-0.1){30}{\line(1,0){0.1}}
\epi
&
\raisebox{10pt}{
$\displaystyle\frac{-2g}{(4\pi)^2\ep}
(2\la+3\xi g^2-3g^2)\vp k_\mu$
}
\\
\hline
\bpi(48,30)(-8,-10)
\put(-9,0){$\mu$}
\put(34,0){$\nu$}
\multiput(2,3)(8,0){4}{\oval(4,4)[t]}
\multiput(6,3)(8,0){4}{\oval(4,4)[b]}
\put(16,3){\circle*{3}}
\put(16,3){\circle{5}}
\epi
&
\raisebox{10pt}{
$\displaystyle
\frac{i}{(4\pi)^2\ep}\left[\left(
\frac{2}{3}g^2k^2+2(3+\xi)g^4\vp^2\right)
g_{\mu\nu}-\frac{8}{3}g^2k_\mu k_\nu\right]$
}
\\
\hline
\bpi(32,30)(0,-10)
\multiput(0,3)(4,0){9}{\circle*{1}}
\put(16,3){\circle*{3}}
\put(16,3){\circle{5}}
\multiput(6.5,3)(-0.1,0.1){30}{\line(1,0){0.1}}
\multiput(6.5,3)(-0.1,-0.1){30}{\line(1,0){0.1}}
\multiput(26.5,3)(-0.1,0.1){30}{\line(1,0){0.1}}
\multiput(26.5,3)(-0.1,-0.1){30}{\line(1,0){0.1}}
\epi
&
\raisebox{10pt}{
$\displaystyle
\frac{-i}{(4\pi)^2\ep}
(4\la\xi g^2+6\xi^2g^4)\vp^2$
}
\\
\end{tabular}
\protect\vsa\\
\begin{tabular}{c|c||c|c}
\bpi(32,40)
\put(0,0){\line(1,1){16}}
\put(16,16){\line(1,-1){16}}
\put(16,16){\line(0,1){20}}
\put(16,16){\circle*{3}}
\put(16,16){\circle{5}}
\epi
&\raisebox{12pt}{$\displaystyle\frac{-12i}{(4\pi)^2\ep}
(10\la^2+2\la\xi g^2+3g^4)\vp$}
&
\bpi(48,40)(-8,-2)
\multiput(0,2)(4,4){4}{\oval(4,4)[br]}
\multiput(4,2)(4,4){4}{\oval(4,4)[tl]}
\multiput(16,14)(4,-4){4}{\oval(4,4)[tr]}
\multiput(20,14)(4,-4){4}{\oval(4,4)[bl]}
\put(16,16){\line(0,1){20}}
\put(16,16){\circle*{3}}
\put(16,16){\circle{5}}
\put(-8,-2){$\mu$}
\put(34,-2){$\nu$}
\epi
&\raisebox{12pt}{$\displaystyle\frac{4i}{(4\pi)^2\ep}
(3+\xi)g^4\vp g_{\mu\nu}$}
\\
\hline
\bpi(32,28)
\put(0,0){\line(1,1){32}}
\put(0,32){\line(1,-1){32}}
\put(16,16){\circle*{3}}
\put(16,16){\circle{5}}
\epi
&\raisebox{12pt}{$\displaystyle\frac{-12i}{(4\pi)^2\ep}
(10\la^2+2\la\xi g^2+3g^4)$}
&
\bpi(48,36)(-8,-2)
\multiput(0,2)(4,4){4}{\oval(4,4)[br]}
\multiput(4,2)(4,4){4}{\oval(4,4)[tl]}
\multiput(16,14)(4,-4){4}{\oval(4,4)[tr]}
\multiput(20,14)(4,-4){4}{\oval(4,4)[bl]}
\put(16,16){\line(1,1){16}}
\put(0,32){\line(1,-1){16}}
\put(16,16){\circle*{3}}
\put(16,16){\circle{5}}
\put(-8,-2){$\mu$}
\put(34,-2){$\nu$}
\epi
&\raisebox{12pt}{$\displaystyle\frac{4i}{(4\pi)^2\ep}
(3+\xi)g^4g_{\mu\nu}$}
\\
\hline
\bpi(32,34)
\multiput(0,0)(0.25,0.25){12}{\line(1,0){0.1}}
\multiput(6,6)(0.25,0.25){12}{\line(1,0){0.1}}
\multiput(12,12)(0.25,0.25){12}{\line(1,0){0.1}}
\multiput(17,17)(0.25,0.25){12}{\line(1,0){0.1}}
\multiput(23,23)(0.25,0.25){12}{\line(1,0){0.1}}
\multiput(29,29)(0.25,0.25){12}{\line(1,0){0.1}}
\multiput(0,32)(0.25,-0.25){12}{\line(1,0){0.1}}
\multiput(6,26)(0.25,-0.25){12}{\line(1,0){0.1}}
\multiput(12,20)(0.25,-0.25){12}{\line(1,0){0.1}}
\multiput(17,15)(0.25,-0.25){12}{\line(1,0){0.1}}
\multiput(23,9)(0.25,-0.25){12}{\line(1,0){0.1}}
\multiput(29,3)(0.25,-0.25){12}{\line(1,0){0.1}}
\put(16,16){\circle*{3}}
\put(16,16){\circle{5}}
\epi
&\raisebox{12pt}{$\displaystyle\frac{-12i}{(4\pi)^2\ep}
(10\la^2-6\la\xi g^2+3g^4)$}
&
\bpi(48,38)(-8,-2)
\multiput(0,2)(4,4){4}{\oval(4,4)[br]}
\multiput(4,2)(4,4){4}{\oval(4,4)[tl]}
\multiput(16,14)(4,-4){4}{\oval(4,4)[tr]}
\multiput(20,14)(4,-4){4}{\oval(4,4)[bl]}
\multiput(17,17)(0.25,0.25){12}{\line(1,0){0.1}}
\multiput(23,23)(0.25,0.25){12}{\line(1,0){0.1}}
\multiput(29,29)(0.25,0.25){12}{\line(1,0){0.1}}
\multiput(0,32)(0.25,-0.25){12}{\line(1,0){0.1}}
\multiput(6,26)(0.25,-0.25){12}{\line(1,0){0.1}}
\multiput(12,20)(0.25,-0.25){12}{\line(1,0){0.1}}
\put(16,16){\circle*{3}}
\put(16,16){\circle{5}}
\put(-8,-2){$\mu$}
\put(34,-2){$\nu$}
\epi
&\raisebox{12pt}{$\displaystyle\frac{12i}{(4\pi)^2\ep}
(1-\xi)g^4g_{\mu\nu}$}
\\
\hline
\bpi(32,40)
\multiput(0,0)(0.25,0.25){12}{\line(1,0){0.1}}
\multiput(6,6)(0.25,0.25){12}{\line(1,0){0.1}}
\multiput(12,12)(0.25,0.25){12}{\line(1,0){0.1}}
\multiput(17,15)(0.25,-0.25){12}{\line(1,0){0.1}}
\multiput(23,9)(0.25,-0.25){12}{\line(1,0){0.1}}
\multiput(29,3)(0.25,-0.25){12}{\line(1,0){0.1}}
\put(16,16){\line(0,1){20}}
\put(16,16){\circle*{3}}
\put(16,16){\circle{5}}
\epi
&\raisebox{14pt}{
$\begin{array}{rl}
\displaystyle\frac{-4i}{(4\pi)^2\ep}
(\!\!\!\!\!&10\la^2+2\la\xi g^2\\
&+3g^4+3\xi^2g^4)\vp
\end{array}$}
&
\bpi(32,40)
\multiput(0,0)(3,3){6}{\circle*{1}}
\multiput(32,0)(-3,3){6}{\circle*{1}}
\put(16,16){\line(0,1){20}}
\put(23.5,4.5){\line(1,0){4}}
\put(27.5,4.5){\line(0,1){4}}
\put(3.5,7.5){\line(1,0){4}}
\put(7.5,3.5){\line(0,1){4}}
\put(16,16){\circle*{3}}
\put(16,16){\circle{5}}
\epi
&\raisebox{12pt}{$\displaystyle\frac{-4i}{(4\pi)^2\ep}
(2\la\xi g^2+3\xi^2g^4)\vp$}
\\
\hline
\bpi(32,38)
\multiput(0,0)(0.25,0.25){12}{\line(1,0){0.1}}
\multiput(6,6)(0.25,0.25){12}{\line(1,0){0.1}}
\multiput(12,12)(0.25,0.25){12}{\line(1,0){0.1}}
\multiput(17,15)(0.25,-0.25){12}{\line(1,0){0.1}}
\multiput(23,9)(0.25,-0.25){12}{\line(1,0){0.1}}
\multiput(29,3)(0.25,-0.25){12}{\line(1,0){0.1}}
\put(16,16){\line(1,1){16}}
\put(0,32){\line(1,-1){16}}
\put(16,16){\circle*{3}}
\put(16,16){\circle{5}}
\epi
&\raisebox{14pt}{
$\begin{array}{rl}
\displaystyle\frac{-4i}{(4\pi)^2\ep}
(\!\!\!\!\!&10\la^2+2\la\xi g^2\\
&+3g^4+3\xi^2g^4)
\end{array}$}
&
\bpi(32,38)
\multiput(0,0)(3,3){6}{\circle*{1}}
\multiput(32,0)(-3,3){6}{\circle*{1}}
\put(16,16){\line(1,1){16}}
\put(0,32){\line(1,-1){16}}
\put(23.5,4.5){\line(1,0){4}}
\put(27.5,4.5){\line(0,1){4}}
\put(3.5,7.5){\line(1,0){4}}
\put(7.5,3.5){\line(0,1){4}}
\put(16,16){\circle*{3}}
\put(16,16){\circle{5}}
\epi
&\raisebox{12pt}{$\displaystyle\frac{-4i}{(4\pi)^2\ep}
(2\la\xi g^2+3\xi^2g^4)$}
\\
\hline
\bpi(36,42)
\put(0,0){\line(1,1){16}}
\multiput(17,15)(0.25,-0.25){12}{\line(1,0){0.1}}
\multiput(23,9)(0.25,-0.25){12}{\line(1,0){0.1}}
\multiput(29,3)(0.25,-0.25){12}{\line(1,0){0.1}}
\multiput(16,18)(0,8){3}{\oval(4,4)[l]}
\multiput(16,22)(0,8){2}{\oval(4,4)[r]}
\put(22,6){\line(1,0){4}}
\put(26,6){\line(0,1){4}}
\put(5,9){\line(1,0){4}}
\put(9,5){\line(0,1){4}}
\put(16,16){\circle*{3}}
\put(16,16){\circle{5}}
\put(-6,9){$k_1$}
\put(28,9){$k_2$}
\put(20,34){$\mu$}
\epi
&\raisebox{14pt}{
$\!\!\!\!\begin{array}{rl}
\displaystyle\frac{2g}{(4\pi)^2\ep}
[\!\!\!\!\!&(2\la+\xi g^2+3g^2)k_{1\mu}
\\&-(2\la+3\xi g^2-3g^2)k_{2\mu}]
\end{array}\!\!\!$}
&
\bpi(32,40)
\multiput(17,17)(0.25,0.25){12}{\line(1,0){0.1}}
\multiput(23,23)(0.25,0.25){12}{\line(1,0){0.1}}
\multiput(29,29)(0.25,0.25){12}{\line(1,0){0.1}}
\multiput(0,32)(0.25,-0.25){12}{\line(1,0){0.1}}
\multiput(6,26)(0.25,-0.25){12}{\line(1,0){0.1}}
\multiput(12,20)(0.25,-0.25){12}{\line(1,0){0.1}}
\put(16,16){\circle*{3}}
\put(16,16){\circle{5}}
\multiput(0,0)(3,3){6}{\circle*{1}}
\multiput(32,0)(-3,3){6}{\circle*{1}}
\put(23.5,4.5){\line(1,0){4}}
\put(27.5,4.5){\line(0,1){4}}
\put(3.5,7.5){\line(1,0){4}}
\put(7.5,3.5){\line(0,1){4}}
\epi
&\raisebox{12pt}{$\displaystyle\frac{4i}{(4\pi)^2\ep}
(2\la\xi g^2-\xi^2g^4)$}
\end{tabular}
\caption{One-loop counterterms in \rxb gauge for the Abelian
Higgs model for $\sg_R=1$. $k_\mu$ is the momentum flowing through the
two-point functions.}
\label{counterterms1}
\end{table}

\section*{Appendix B}
The $Z_x$ for the SU(2)-Higgs model of section \ref{su2higgs} in
MS are for general $\sg$ at the one-loop level:
\bea
\mu^\ep Z_H &=& 1+\frac{g^2}{\left(4\pi\right)^2\ep}\left(\frac{9}{2}+3\xi
-\frac{3\xi}{2\sg}\right)
\\
\mu^\ep Z_G &=& 1+\frac{g^2}{\left(4\pi\right)^2\ep}\left(\frac{9}{2}-\xi
-\frac{3\xi}{2\sg}\right)
\\
\mu^\ep Z_A &=& 1+\frac{g^2}{\left(4\pi\right)^2\ep}\left(\frac{25}{3}
-\frac{2\xi}{\sg}\right)
\\
\mu^\ep Z_\et &=& 1+\frac{g^2}{\left(4\pi\right)^2\ep}
\left(3-\frac{\xi}{\sg}\right)
\\
Z_m &=& 1+\frac{g^2}{\left(4\pi\right)^2\ep}\left(\frac{12\la}{g^2}
-\frac{9}{2}\right)
\\
\mu^{-\ep}Z_\la &=& 1+\frac{g^2}{\left(4\pi\right)^2\ep}
\left(\frac{24\la}{g^2}-9+\frac{9g^2}{8\la}\right)
\\
\mu^{-\ep}Z_g &=& 1+\frac{g^2}{\left(4\pi\right)^2\ep}
\left(\frac{-43}{3}\right)
\\
Z_\xi &=& 1+\frac{g^2}{\left(4\pi\right)^2\ep}\left(\frac{4\la}{g^2}
+\frac{41}{6}-\xi+\frac{\xi }{\sg}\right)\\
Z_\sg &=& 1+\frac{g^2}{\left(4\pi\right)^2\ep}\left(\frac{4\la}{g^2}
-\frac{3}{2}-\xi+4\al\xi+\frac{3\xi}{\sg}+\frac{\sg\xi}{2}
+4\al^2\sg\xi\right)
\\
Z_\al &=& 1+\frac{g^2}{\left(4\pi\right)^2\ep}\left(-\frac{4\la}{g^2}
+\frac{3}{2}+\xi+\frac{\xi}{4\al}+2\al\xi-\frac{\xi}{\sg}\right)
\eea
In an actual application to compute the EP we would set $\sg=1$.

\section*{Appendix C}
The $Z_x$ for the SO(3)-Higgs model of section \ref{so3higgs} in
MS are for general $\sa$ and $\sw$ at the one-loop level:
\bea
\mu^\ep Z_H &=& 1+\frac{g^2}{(4\pi)^2\ep}\left(12+8\xi-\frac{4\xi}{\sw}\right)
\\
\mu^\ep Z_G &=& 1+\frac{g^2}{(4\pi)^2\ep}\left(12-4\xi-\frac{2\xi}{\sa}
-\frac{2\xi}{\sw}\right)
\\
\mu^\ep Z_A &=& 1+\frac{g^2}{(4\pi)^2\ep}\left(8+6\ga\xi-\frac{2\xi}{\sw}
+\frac{2\ga\xi^2}{\sw}\right)
\\
\mu^\ep Z_W &=& 1+\frac{g^2}{(4\pi)^2\ep}\left(8-3\ga\xi-\frac{\xi}{\sa}
-\frac{\xi}{\sw}-\frac{\ga\xi^2}{\sa}\right)
\\
\mu^\ep Z_{\eb_A}Z_{\et_A} &=& 1+\frac{g^2}{(4\pi)^2\ep}
\left(3-2\al\xi+2\be\xi-3\ga\xi-\frac{\xi}{\sw}-2\be\ga\xi^2
-\frac{\ga\xi^2}{\sw}\right)
\\
\mu^\ep Z_{\eb_W}Z_{\et_W} &=& 1+\frac{g^2}{(4\pi)^2\ep}\left(3+\al\xi
-\be\xi+\frac{3\ga\xi}{2}-\frac{\xi}{2\sa}-\frac{\xi}{2\sw}
+\be\ga\xi^2+\frac{3\ga^2\xi^2}{2}
\right.
\nn&&
\left.
-\frac{\ga\xi^2}{\sa}+\frac{3\ga\xi^2}{2\sw}
-\frac{\ga^2\xi^3}{2\sa}\right)
\\
\mu^\ep Z_{\eb_A}Z_{\et_W} &=& 1+\frac{g^2}{(4\pi)^2\ep}\left(3-3\ga\xi
-\frac{\xi}{\sa}-2\be\ga\xi^2-\frac{\ga\xi^2}{\sw}\right)
\\
Z_m &=& 1+\frac{g^2}{(4\pi)^2\ep}\left(\frac{10\la}{g^2}-12\right)
\\
\mu^{-\ep}Z_\la &=& 1+\frac{g^2}{(4\pi)^2\ep}\left(\frac{22\la}{g^2}-24
+\frac{12g^2}{\la}\right)
\\
\mu^{-\ep}Z_g &=& 1+\frac{g^2}{(4\pi)^2\ep}\left(-14\right)
\\
Z_\xi &=& 1+\frac{g^2}{(4\pi)^2\ep}\left(\frac{4\la}{g^2}-1-6\ga-2\xi-\al\xi
+\be\xi-\frac{3\ga\xi}{2}+\frac{\xi}{2\sa}+\frac{9\xi}{2\sw}\right.
\nn&&
\left.-\be\ga\xi^2-\frac{3\ga^2\xi^2}{2}-\frac{\ga\xi^2}{\sa}
-\frac{3\ga\xi^2}{2\sw}+\frac{\ga^2\xi^3}{2\sa}\right)
\\
Z_{\sa} &=& 1+\frac{g^2}{(4\pi)^2\ep}\left(\frac{4\la}{g^2}-9-6\ga-2\xi
-\al\xi+5\be\xi-\frac{15\ga\xi}{2}+\frac{\xi}{2\sa}\right.
\nn&&
\left.+4\be^2\sa\xi+\frac{13\xi}{2\sw}-5\be\ga\xi^2-\frac{3\ga^2\xi^2}{2}
-\frac{\ga\xi^2}{\sa}-\frac{7\ga\xi^2}{2\sw}+\frac{\ga^2\xi^3}{2\sa}\right)
\\
Z_{\sw} &=& 1+\frac{g^2}{(4\pi)^2\ep}\left(\frac{4\la}{g^2}-9-6\ga-2\xi
+\al\xi+3\be\xi+\frac{3\ga\xi}{2}+\frac{3\xi}{2\sa}+\frac{11\xi}{2\sw}\right.
\nn&&
+2\sw\xi+4\al\be\sw\xi+6\ga^2\sw\xi+\be\ga\xi^2+\frac{3\ga^2\xi^2}{2}
-\frac{3\ga\xi^2}{\sa}+\frac{3\ga\xi^2}{2\sw}
\nn&&
\left.+\frac{3\ga^2\xi^3}{2\sa}\right)
\\
Z_\al &=& 1+\frac{g^2}{(4\pi)^2\ep}\left(-\frac{4\la}{g^2}+9+6\ga+2\xi
+\frac{2\xi}{\al}+\al\xi+\be\xi-\frac{3\ga\xi}{2}+\frac{6\ga^2\xi}{\al}\right.
\nn&&
-\frac{3\xi}{2\sa}-\frac{7\xi}{2\sw}+3\be\ga\xi^2-\frac{3\ga^2\xi^2}{2}
+\frac{3\ga\xi^2}{\sa}-\frac{3\ga\xi^2}{2\sw}+\frac{4\be\ga\xi^2}{\al\sw}
\nn&&
\left.-\frac{3\ga^2\xi^3}{2\sa}\right)
\\
Z_\be &=& 1+\frac{g^2}{(4\pi)^2\ep}\left(-\frac{4\la}{g^2}+9+6\ga+2\xi
+2\al\xi+3\ga\xi-\frac{5\xi}{\sw}+\frac{\ga\xi^2}{\sw}\right)
\\
Z_\ga &=& 1+\frac{g^2}{(4\pi)^2\ep}\left(-\frac{4\la}{g^2}+\frac{11}{2}
+\frac{\al}{2\ga}-\frac{\be}{2\ga}+6\ga+2\xi+\frac{\al\xi}{2}
-\frac{3\ga\xi}{2}\right.
\nn&&
\left.-\frac{\xi}{2\sa}-\frac{2\xi}{\sw}+\frac{\be\ga\xi^2}{2}
+\frac{\ga\xi^2}{\sa}-\frac{\ga\xi^2}{\sw}-\frac{\ga^2\xi^3}{2\sa}\right).
\eea
In an actual application to compute the EP we would set $\sw=1$. Notice
that because of ghost number conservation (i.e.\ $\eb_x$ and $\et_y$
appear always in pairs $\eb_x\et_y$) the $Z_{\eb_W}$, $Z_{\et_W}$,
$Z_{\eb_A}$ and $Z_{\et_A}$ are not determined uniquely, but only the
combinations $Z_{\eb_A}Z_{\et_A}$, $Z_{\eb_W}Z_{\et_W}$, $Z_{\eb_A}Z_{\et_W}$
and $Z_{\eb_W}Z_{\et_A}$, of which only three are independent.

\setlength{\baselineskip}{14pt}

\end{document}